\documentclass[twocolumn,trackchanges]{aastex701}

\usepackage{orcidlink}
\usepackage[version=4]{mhchem}
\usepackage{mathtools}
\usepackage{placeins}
\usepackage{float}

\begin{document}

\title{The Effects of Non-ideal Mixing in Planetary Magma Oceans and Atmospheres}

\shortauthors{Werlen et al.}
\correspondingauthor{Aaron Werlen}
\shorttitle{}

\author[orcid=0009-0005-1133-7586, sname='Werlen']{Aaron Werlen}
\affiliation{Institute for Particle Physics and Astrophysics, ETH Zurich, CH-8093 Zurich, Switzerland}
\email[show]{awerlen@ethz.ch}  

\author[orcid=0000-0002-1299-0801, sname='Young']{Edward D. Young}
\affiliation{Department of Earth, Planetary, and Space Sciences, University of California, Los Angeles, CA 90095, USA}
\email{eyoung@epss.ucla.edu}

\author[orcid=0000-0002-0298-8089, sname='Schlichting']{Hilke E. Schlichting}
\affiliation{Department of Earth, Planetary, and Space Sciences, University of California, Los Angeles, CA 90095, USA}
\email{hilke@ucla.edu}

\author[orcid=0000-0001-6110-4610, sname='Dorn']{Caroline Dorn}
\affiliation{Institute for Particle Physics and Astrophysics, ETH Zurich, CH-8093 Zurich, Switzerland}
\email{dornc@ethz.ch}

\author[orcid=0000-0002-0794-2717, sname='Shahar']{Anat Shahar}
\affiliation{Earth and Planets Laboratory, Carnegie Institution for Science, Washington DC, USA.}
\email{ashahar@carnegiescience.edu}

\begin{abstract}
Sub-Neptunes with hydrogen-rich envelopes are expected to sustain long-lived magma oceans that continuously exchange volatiles with their overlying atmospheres. Capturing these interactions is key to understanding the chemical evolution and present-day diversity of sub-Neptunes, super-Earths, and terrestrial planets, particularly in light of new JWST observations and upcoming missions. Recent advances in both geochemistry and astrophysics now allow the integration of experimental constraints and thermodynamic models across melt, metal, and gas phases. Here we extend a global chemical equilibrium model to include non-ideal behavior in all three phases. Our framework combines fugacity corrections for gas species with activity coefficients for silicate and metal species, enabling a fully coupled description of volatile partitioning. We show that for planetary embryos (0.5 M$_\oplus$ at 2350~K), non-ideality introduces only modest corrections to atmosphere–magma ocean interface (AMOI) pressures, volatile inventories, and interior compositions. In contrast, for sub-Neptunes with higher temperatures ($\approx$ 3000 K) and pressures, non-ideal effects are more pronounced, though still modest in absolute terms—typically within 20\% and at most a factor of two. Including activity and fugacity coefficients simultaneously increases the AMOI pressure, enhances water retention in the mantle and the envelope. Our results demonstrate that non-ideality must be treated globally: applying corrections to only one phase leads to incomplete or even misleading trends. These findings highlight the importance of self-consistent global thermodynamic treatments for interpreting atmospheric spectra and interior structures of sub-Neptunes and super-Earths.
\end{abstract}



\section{Introduction}\label{sec:intro}

Over the past decade, the detection and characterization of exoplanets, particularly sub-Neptunes, has increased drastically \citep[e.g.,][]{madhusudhan_carbon-bearing_2023, benneke_jwst_2024, felix_competing_2025, madhusudhan_new_2025}. With these advances, the role of interior-atmosphere interactions in shaping atmospheric composition has become a central question.  

Sub-Neptunes with hydrogen-rich envelopes of even only a few wt\% can sustain surface temperatures of several thousand Kelvin over gigayear timescales, resulting in long-lived magma oceans \citep{ginzburg_super-earth_2016, misener_importance_2022}. These magma oceans must interact chemically with their overlying gaseous envelopes, thereby modifying their composition \citep[e.g.,][]{ginzburg_super-earth_2016, chachan_role_2018, kite_superabundance_2019, kite_water_2021, schlichting_chemical_2022, misener_importance_2022, misener_atmospheres_2023, seo_role_2024, burn_water-rich_2024, werlen_atmospheric_2025, werlen_sub-neptunes_2025, nixon_magma_2025, lichtenberg_constraining_2025}.  These modifications may be manifest in observable chemical species in the atmospheres \citep[e.g.,][]{shorttle_distinguishing_2024, werlen_atmospheric_2025,lee_mineral_2025,nixon_magma_2025}.

For a first-order understanding of the effects of magma oceans on observable atmospheres, equilibrium chemical thermodynamics can be assumed \citep[e.g.,][]{schlichting_chemical_2022}. A central question in this endeavor is the fidelity of the thermodynamics used to characterize envelope-melt exchange. The precise values for the thermodynamic parameters necessary to characterize chemical equilibrium are often unknown because of the extreme temperatures and pressures that are relevant compared with terrestrial settings where most thermochemical data are calibrated. In particular, non-ideality of mixing and non-ideal equations of state of even pure phases may well have a profound effect on our understanding of sub-Neptunes.  Here we investigate the effects of non-ideality on the chemical interaction between magma oceans and overlying hydrogen-rich primary atmospheres.  We use the proto-Earth as a fiducial reference case, and then extend our analysis to sub-Neptunes. We do not address in this work the ramifications of complete miscibility between hydrogen-rich envelopes and magma oceans \citep{young_phase_2024,rogers_redefining_2025}, but this is an important aspect of non-ideality for sub-Neptunes.  

This study is structured as follows. In Section~\ref{sec:methods}, we describe the chemical thermodynamics framework, including the treatment of non-ideality in gas, silicate melt, and metal liquid species. Building on this framework, Section~\ref{sec:Results} presents results for Earth and for a fiducial sub-Neptune. In Section~\ref{sec:Discussion}, we discuss the implications of our findings and compare them with other studies that account for non-ideal mixing. Finally, Section~\ref{sec:Conclusions} provides a summary of our main results.

\section{Methods}\label{sec:methods}

\subsection{Chemical Thermodynamics}\label{sec:chemical_network}
We adopt the global chemical equilibrium framework of \citet{schlichting_chemical_2022} as a starting point. The network comprises 18 independent reactions among 25 phase components in coexisting metal, silicate, and gas phases (see Appendix~\ref{app:chemical_network} for more details). Compositions of the phases and their relative abundances are obtained by solving simultaneously for (i) chemical equilibrium, (ii) elemental conservation, and (iii) summing constraints on mole fractions of all phase components for each phase. Our implementation follows that of \citet{schlichting_chemical_2022} with performance improvements detailed in Grimm et al. (in preparation). For the explicit governing equations, see \citet{schlichting_chemical_2022} and the appendices of \citet{young_earth_2023} and \citet{werlen_sub-neptunes_2025}. In prior applications, use of non-ideality was sparing, confined mainly to light elements in metal.  In this study we modify the equations to account for non-ideal behavior in selected, critical gas and silicate species, as well as in metal (Section~\ref{sec:nonideality}). Furthermore, Appendix~\ref{app:equilibrium_methods} provides a concise overview of the solution methodology, clarifies notation, and demonstrates the equivalence between our Gibbs free-energy minimization approach and the ``extended law of mass action" as described in other studies like \citet{bower_diversity_2025}.

\subsection{Non-Ideality}\label{sec:nonideality}
For each reaction $r$ with stoichiometric coefficients $\nu_i$ (positive for products, negative for reactants), the equilibrium condition is

\begin{equation}\label{eq:eq_cond}
    \sum_i \nu_i \mu_i = 0,
\end{equation}
where the chemical potentials $\mu_i$ for a species $i$ in the host phases are given by

\begin{equation}\label{eq:conditions}
    \mu_i =
    \begin{cases}
      \Delta \hat{G}_{f,i}^{T, P \lor P^\circ, *} + RT \ln a_i & \text{metal/silicate species},\\[6pt]
      \Delta \hat{G}_{f,i}^{T, P^\circ, *} + RT \ln\!\left(\dfrac{f_i}{P^\circ}\right) & \text{gas species},
    \end{cases}
\end{equation}

\noindent where $a_i=\gamma_i x_i$ corresponds to the activity, $x_i$ is the mole fraction of species $i$, $\gamma_i$ is the activity coefficient that may or may not be pressure dependent, $f_i=\phi_i x_i P$ is the fugacity specified by the fugacity coefficient $\phi_i$ for the impure species, mole fraction $x_i$ for species $i$ in the gas or envelope phase, and atmosphere-magma ocean interface (AMOI) pressure $P$, $R$ is the gas constant, and $T$ the temperature evaluated at either the AMOI or the silicate-metal interface (SMI). In Equation \ref{eq:conditions}, two standard states are invoked. The standard-state chemical potentials for species in the condensed phases are the molar Gibbs free energies of formation for pure species $i$ (purity signified by the $^*$ superscript), $\Delta \hat{G}_{f,i}^{T, P \lor P^\circ, *}$ at $T$ and at either $P$ or $P^\circ$.  The standard-state chemical potentials for the gaseous, or supercritical envelope, species are similar to the Gibbs free energies of formation for the pure species at $T$, but at a reference pressure $P^\circ$ =~1~bar, or $\Delta \hat{G}_{f,i}^{T, P^\circ, *}$.  In these equations, the activity coefficients $\gamma_i$ correct the standard-state chemical potentials (molar free energies of formation) for both composition, and where appropriate, pressure. The fugacity coefficients $\phi_i$ correct for both the non-ideal equation of state behavior for the pure gas species and the non-ideality of mixing, if known (i.e., $\phi_i = \gamma_i \phi_i^*$ if $\gamma_i$ corrects for composition).   

In this study we examine the non-ideality for the gaseous species \ce{H2}, \ce{H2O}, \ce{CH4}, \ce{CO2}, and \ce{CO}, the  metal liquid  species \ce{O}, \ce{Si} and \ce{H}, and the silicate melt species \ce{H2O} and \ce{H2}. For all other species we set $\gamma_i=\phi_i=1$.  We focus on the non-ideality of H$_2$ and H$_2$O in both gas and melt as these species are of particular importance in planetary applications, and because the equations of state of these species at the high $P$ and $T$ conditions of interest are well characterized.

\subsubsection{Fugacity Coefficients for Gas Species}
\begin{figure*}[t]
    \centering
    \includegraphics[width=1\textwidth]{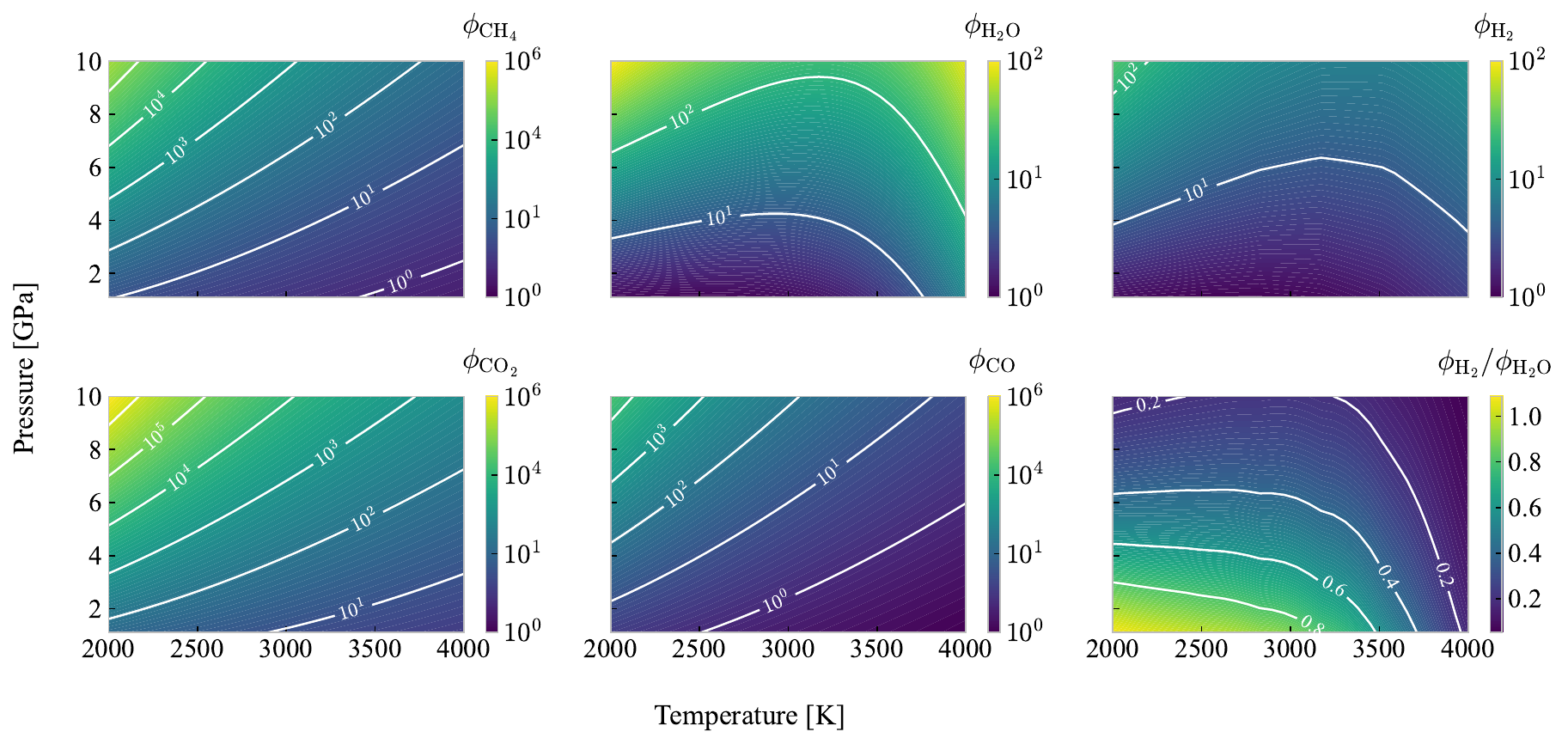}
    \caption{Contour plots of the fugacity coefficients ($\phi_i$) for \ce{H2O}, \ce{CH4}, \ce{CO2}, \ce{CO}, \ce{H2}, and for the ratio $\phi_{\ce{H2}}/\phi_{\ce{H2O}}$, shown as a function of pressure and temperature. A value of $\phi_i=1$ corresponds to ideal behavior, while deviations indicate non-ideality in the gas phase. For direct comparison, the color scale is shared across \ce{CH4}, \ce{CO2}, and \ce{CO}, and separately for \ce{H2O} and \ce{H2}. Fugacity coefficients become large at high pressures, but remain relatively small over the pressure range relevant for sub-Neptunes, which typically have surface pressures below 10~GPa \citep{young_phase_2024,gilmore_core-envelope_2026}}
        
    \label{fig:gamma_all_species}
\end{figure*}

The equation of state (EoS) for \ce{H2} is obtained by interpolating the tabulated values of $\rho$ from \citet{Chabrier2019} and comparing them with an ideal-gas EoS. The tabulated data span pressures from $10^{-9}$ to $10^{13}$~GPa and temperatures from $10^{2}$ to $10^{8}$~K. The fugacity coefficient of \ce{H2}, $\phi^*$, is then obtained from the compressibility factor, $Z(P,T) = \rho_{\rm ideal}/\rho$, using 

\begin{equation}
\ln(\phi^*)=\int_0^{P}\frac{Z(P,T)-1}{P}dP,
\label{eqn:z}
\end{equation}
where $\rho_{\rm ideal}$ is the density of the ideal gas and $\rho$ is the density at $P$ and $T$ of the real gas.  Results are shown in Figure~\ref{fig:gamma_all_species}.

A similar approach is used to derive $\phi^*$ for H$_2$O. \cite{Haldemann} provide a compilation of a continuous EoS for water from $10^{-10}$~GPa to 400~TPa and 150 to $10^5$~K. We use this compilation to extract values for $\phi(P,T)$ and apply Equation \ref{eqn:z} to derive fugacity coefficients.  ~

Fugacity coefficients of even pure species can be challenging to evaluate, especially where temperatures and pressures are more extreme than those encountered in Earth's crust (where much of the work on fugacity coefficients has concentrated). Corresponding states equations, in which functional forms for compressibility $Z_i$, and thus the integral  $\int[(Z_i-1)/P] dP = \ln(\phi_i)$, are expressed in terms of the critical temperatures and pressures for each gas, $T_c$ and $P_c$, take forms similar to

\begin{equation}
Z = \frac{P\hat{V}}{RT} = A(T_{\rm c}) + B(T_{\rm c})P_{\rm c} + C(T_{\rm c})P^2_{\rm c} + D(T_{\rm c})P^3_{\rm c}.
\end{equation} 
where $\hat{V}$ is the molar volume of the gas.  These formulations work reasonably well at moderately high $T$ and $P$ for relatively inert species like CO, CO$_2$, and to a lesser degree CH$_4$.  They fail for H$_2$ at higher $T$ and $P$, requiring instead an explicit equation of state, as described above. In what follows, we evaluate fugacity coefficients for CO, CO$_2$, and CH$_4$ based on the corresponding states equations of \cite{Shi_Saxena_1992} (their table 1). We caution, however, that the applicable range for these coefficients is for maximum $P$ and $T$ of about 2~GPa and 2500~K, requiring a significant extrapolation  to sub-Neptune surface conditions (Figure \ref{fig:gamma_all_species}). We do not include non-ideal interaction parameters (e.g., activity coefficients based on excess free energies) for the fugacities, in part because multi-component interaction parameters for \ce{H2}-rich compositions, as opposed to aqueous systems \citep[e.g.,][]{Kerrick_Jacobs_1981}, are largely unknown, to our knowledge.  

\subsubsection{Activity Coefficients for Silicate Species}

An accurate portrayal of the chemistry of melts can require characterizing non-ideal  activities for chemical constituents. This in turn requires a vast amount of thermodynamic data, and can be plagued by kinetic factors \citep{Walker2022}. For example, \cite{Ford1983} used activities to characterize igneous processes on Earth and noted that ``the derivation of component activities from phase compositions is a fundamental problem in equilibrium thermodynamics." Codes such as the MAGMA code \citep{schaefer_outgassing_2007, schaefer_chemistry_2009} make use of laboratory measurements of equilibrium vapor pressures to constrain activities of simple oxides in melts, while other models like pMELTS \citep{Ghiorso2002}, make use of interpolation between experimental data to derive regular or sub-regular solution models involving more complex melt species (e.g., Mg$_2$SiO$_4$ rather than just SiO$_2$ and MgO). 

The thermodynamic activities of components in silicate and oxide melts  are most often "measured" by comparing the compositions of coexisting crystals and melts to  equilibrium constants derived from existing thermodynamic data for melts  \citep{Wood2013}.  For example, the reaction 
\setlength{\abovedisplayskip}{3pt}
\begin{equation}
    {\rm Fe_{solid}} + 1/2 {\rm O_2} = {\rm FeO_{melt}}
\label{eq:FeO}
\end{equation}
is described by the equilibrium constant
\setlength{\abovedisplayskip}{3pt}
\begin{equation}
    k_{\rm eq}=\frac{a_{\rm FeO}}{a_{\rm Fe} P_{\rm O_2}^{1/2}{}}=\exp{\left(\frac{-\Delta \hat{G}^0}{RT}\right)},
\label{eq:Keq_FeO}
\end{equation}

\noindent where $a_{\rm FeO}$ refers to the activity of  FeO in the melt phase, $a_{\rm Fe}$ refers to the activity of Fe in the solid phase, and $\Delta \hat{G}^0$ is the standard-state molar Gibbs free energy for the reaction.  The partial pressure of \ce{O2}, $P_{\rm O_2}$, can be replaced with the fugacity of oxygen, $f_{\rm O_2}$, where mixing in the vapor phase is non-ideal. Recalling that activity of a component $i$ is the effective concentration given by $a_i = \gamma_i x_i$ where $x_i$ is mole fraction (concentration) of the component in the melt, and $\gamma_i$ is the activity coefficient, departures from ideal mixing are expressed as deviations in $\gamma_i$ from unity.  There is, therefore, usually a non-ideal aspect to activities in melts that must be characterized.

Unlike crystalline materials where mixing of cations occurs on relatively well understood crystallographic sites, making the thermodynamics of mixing tractable, silicate melts consist of polymer units that are not well defined {\it a priori} \citep[e.g.,][]{Hess1971, Mysen1982, Nesbitt2020}. The thermodynamics of impure silicate and oxide melts are therefore less well known. 

The argument of the exponential on the right side of Equation \ref{eq:Keq_FeO} consists of the Gibbs free energy for the reaction for a standard state of pure phases (pure Fe solid, pure FeO melt, and pure O$_2$ vapor).  In order to determine the activity of FeO in the melt in this example, one must have an activity-composition model for the solid (e.g., $\gamma_{\rm Fe}$ and/or a mixing-on-sites activity model), a measure of the oxygen fugacity, and standard-state thermodynamic data for the pure solid, vapor, and melt species. Both experiments and data for natural rocks have been used to obtain activities of melt species using this approach  \citep{Carmichael1970,Nicholls1971,Ryerson1985,Wood2013}.
Thermodynamic data for melts, and to a lesser degree activity models for crystalline solids, present impediments to the accuracy of measured activities in melts, leading to considerable uncertainties.  For example, in the case of \ce{FeO}, \citet{Corgne2008} and \citet{Holzheid1997} suggest $\gamma_{\rm FeO}$ values of $3\pm 1$ for ultramafic melt compositions while \citet{Wood2013} obtain values of $< 1$ for similar melt compositions. 

The identities of components used to characterize the thermodynamics of silicate melts need not be, and indeed cannot be, perfect representations of actual structural components in the melts.  However, if chosen carefully, the components used can facilitate use of ideal mixing, reducing the sensitivity of results on activity coefficients.  In principle, choices of components can be made as a matter of convenience, because the thermodynamic quantities are always relative to a chosen reference state.  It is customary to portray the compositions of melts using simple oxides such as MgO, SiO$_2$, FeO, Na$_2$O, and so forth, as a means of keeping track of oxygens associated with each cation species.  In this practice, all Mg is assigned to MgO, all Fe (Fe$^{2+}$) to FeO, all Na to Na$_2$O, and so on.  Expressing evaporation reactions in terms of these simple oxides has the advantage that the basic thermodynamic entities are more tractable from previous experiments and data bases.  

However, in practice, the thermodynamic activities  of the oxides in melts are often less than or greater than their concentration in the melts.  In the case of \ce{Mg}, as an illustration, the activity of \ce{MgO} is usually found to be much lower than its calculated concentration on the basis of assigning all Mg in the melt to \ce{MgO}.  This is apparently because the coordination of  Mg in silicate melts is fivefold \citep{Henderson2006}, distinct from the sixfold coordination in most silicate crystalline materials. One means of accounting for this behavior is to consider that Mg is bonded to silicate polymers, and not just oxygen as \ce{MgO}. This in turn suggests intra-melt reactions resembling \ce{MgO + SiO2 \rightleftharpoons MgSiO3} take place.  If a set of liquid components representing speciation in the liquids (e.g., \ce{MgSiO3}) can be characterized thermodynamically, the activities of the simple oxides can be accounted for by the speciation assuming ideal mixing among these components. This is the approach employed by the MAGMA code, a tool commonly used in planetary science to obtain estimates of equilibrium vapor pressures for silicate and oxide melts \citep{Fegley1987, Hastie1986}. 
This is also the approach taken by \cite{schlichting_chemical_2022} in their implementation of a global equilibrium model for sub-Neptunes.  

While ideal mixing of well-chosen components in the melt may be a reasonable, if imperfect, assumption for intra-melt reactions among silicate and oxide species, it is becoming clearer that non-ideality dominates the mixing behavior of silicate melts with volatiles, especially \ce{H2}, and should not be ignored. We now know that \ce{MgSiO3} and \ce{H2} become entirely miscible at temperatures and pressures similar to those expected for the interface between sub-Neptune magma oceans and their overlying envelopes \citep{gilmore_core-envelope_2026, young_phase_2024}. A transition from immiscible to entirely miscible species is only possible when enthalpy contributions are considered, which enter through non-ideal mixing.

\begin{figure*}
    \centering
    \includegraphics[width=1\textwidth]{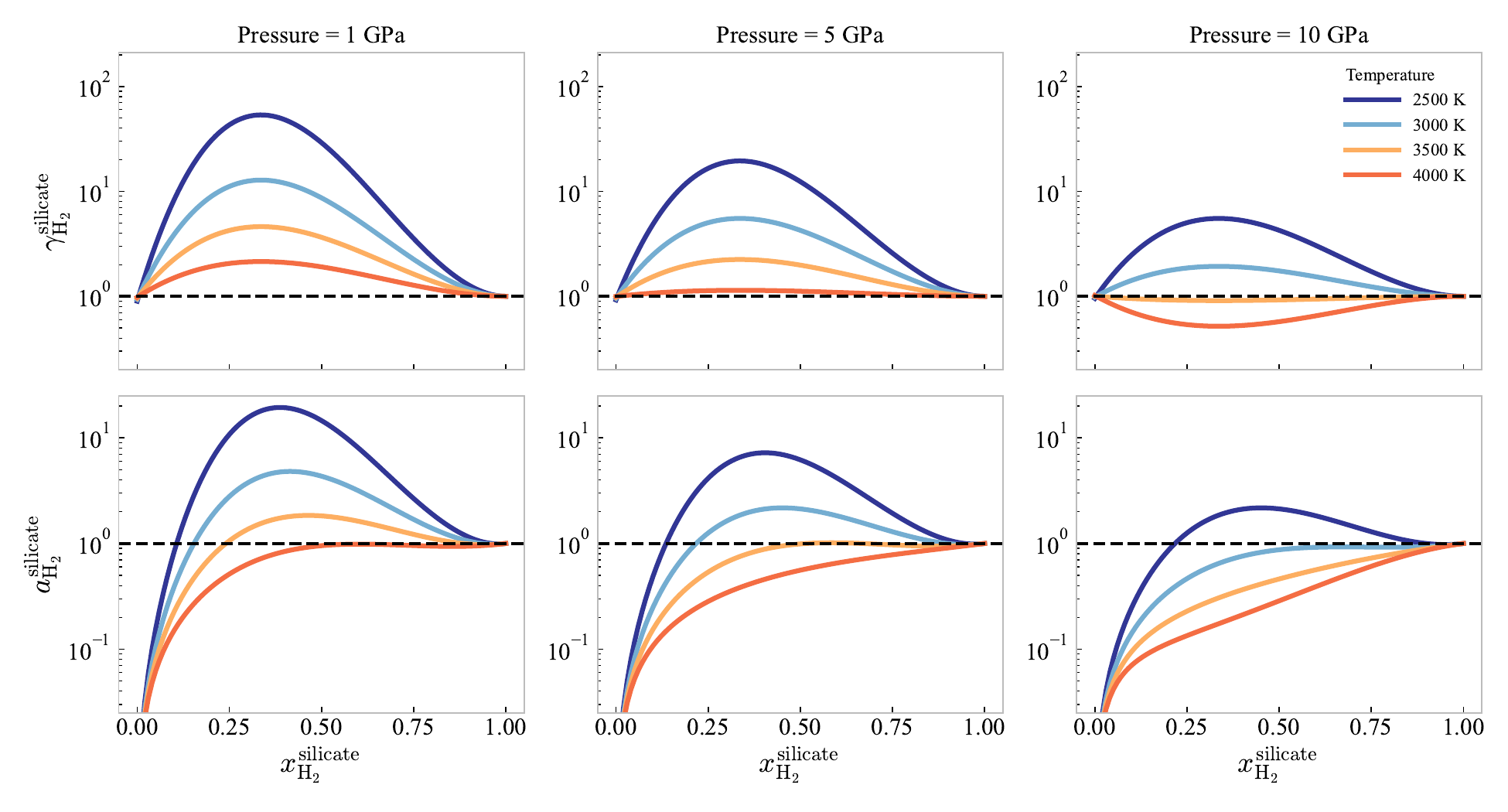}
    \caption{Activity coefficient $\gamma$ and activities $a$ for \ce{H2} in silicate melt as a function of the mole fraction of \ce{H2} in the silicate phase. Curves are shown for different pressures and temperatures. Both the activities and the activity coefficient are strongly temperature dependent. At lower temperatures, the dependence on pressure becomes significant.}
    \label{fig:aH2melt}
\end{figure*}

The sub-regular solution model is often used to characterize non-ideal mixing in melts. We utilize the sub-regular mixing model produced by ab-initio-molecular dynamics (ab-initio-MD) modeling of this system by \cite{gilmore_core-envelope_2026} to specify the activity coefficient for H$_2$ in melts in our models. From $RT\ln(\gamma) = $ $\hat{G}_{\rm Ex} + (1-x_i)\,\partial \hat{G}_{\rm Ex}/\partial x_i$, where $\hat{G}_{\rm Ex}$ is the molar excess Gibbs free energy of mixing and $\partial \hat G_{\mathrm{Ex}}/\partial x_i$ captures its dependence on composition, we arrive at the activity coefficient for silicate \ce{H2} ($\gamma_{\mathrm{H}_2}^{\mathrm{s}}$), consistent with the mixing parameters given by \citet{gilmore_core-envelope_2026}:

\begin{align}\label{eqn:gammaH2_sil}
\ln \gamma_{\mathrm{H_2}}^{\mathrm{s}}
= &\left( 1 - \frac{T}{4,800\,\mathrm{K}}
- \frac{P}{35.0\,\mathrm{GPa}} \right) \\
& \times \frac{(x_{\ce{H2}}^\mathrm{s} - 1)^{2}}{R \, T}
\left( 1,253,900\, x_{\ce{H2}}^\mathrm{s} - 4,950 \right), \nonumber
\end{align}

\noindent where $x_{\rm H_2}^{\rm s}$ is the mole fraction of H$_2$ in the silicate (s) melt. Figure~\ref{fig:aH2melt} shows the activity coefficients $\gamma$ as well as the activities $a$ for \ce{H2} in silicate melt as a function of the mole fraction of \ce{H2} at various pressures and temperatures. The large values of activity at relatively small mole fractions is reminiscent of activity-composition relationships in other silicate-volatile systems \citep[e.g.,][]{Makhluf_2017}. When applying Equation  \ref{eqn:gammaH2_sil} we use the temperature at the AMOI in our calculations.

We are often interested in endogenous production of water in magma oceans \citep{Kite_2020b, schlichting_chemical_2022, young_earth_2023}. In this context, introducing non-ideal activities for H$_2$ in the melt should be balanced with similarly non-ideal activities for H$_2$O in the melt, analogous to the case for hydrogen. \cite{Kovacevic_2022} found that \ce{MgSiO3} and \ce{H2O} are completely miscible at high $T$ and $P$. Taking their model temperature for complete miscibility of about 4000~K at $P\sim 20$~GPa as indicative of the critical (consolute) temperature, $T_{\rm crit}$, for a regular (i.e., symmetric) mixing model, where the interaction parameter that characterizes the non-ideality is $W = 2RT_{\rm crit}$. In a regular solution model for a binary mixture, the molar excess Gibbs free energy of mixing is $\hat G_{\mathrm{Ex}} = W x(1-x)$ where $W$ depicts the magnitude of the non-ideal enthalpy and $x$ is the independent mole fraction composition variable.  It also determines the activity coefficient from the definition of chemical potential in terms of the  excess free energy: $\mu_{i,\mathrm{Ex}} = \partial G_{\mathrm{Ex}}/\partial n_i = RT\ln(\gamma_i)$. We therefore arrive at an expression for the temperature-dependent activity coefficient for water in the melt phase, \ce{H2O} silicate $(\gamma_{\ce{H2O}}^{\mathrm{s}})$ based on this interaction parameter:

\begin{equation}
    \ln(\gamma_{\ce{H2O}}^{\mathrm{s}}) = \frac{74,826}{R \,T} (1-x_{\ce{H2}}^\mathrm{s})^2.
\end{equation}

\noindent This $\gamma_{\ce{H2O}}^{\mathrm{s}}$ is pressure independent, while in reality there is likely a pressure term.

\subsubsection{Activity Coefficients for Metal Species}

We adopt the same activity coefficients for Si metal ($\gamma_{\mathrm{Si}}^{m}$) and O metal ($\gamma_{\mathrm{O}}^{m}$) as defined in \citet{badro_core_2015} and used in \citet{ young_earth_2023} and \citet{werlen_atmospheric_2025,werlen_sub-neptunes_2025}:

\begin{align}
\ln \gamma_{\mathrm{Si}}^{\mathrm{m}}= &-6.65 \frac{1,873}{T_{\mathrm{SMI}}~[\mathrm{K}]} - 12.41 \frac{1,873}{T_{\mathrm{SMI}}~[\mathrm{K}]} \ln(1 - x_{\ce{Si}}^{\mathrm{m}}) \\
&+ 5 \frac{1,873}{T_{\mathrm{SMI}}~[\mathrm{K}]} x_{\ce{O}}^{\mathrm{m}} \left( 1 + \frac{\ln(1 - x_{\ce{O}}^{\mathrm{m}})}{x_{\ce{O}}^{\mathrm{m}}} - \frac{1}{1 - x_{\ce{Si}}^{\mathrm{m}}} \right) \nonumber \\
&- 5 \frac{1,873}{T_{\mathrm{SMI}}~[\mathrm{K}]} (x_{\ce{O}}^{\mathrm{m}})^2  x_{\ce{Si}}^{\mathrm{m}} \left( \frac{1}{1 - x_{\ce{Si}}^{\mathrm{m}}} + \frac{1}{1 - x_{\ce{O}}^{\mathrm{m}}} \right. \nonumber \\
&\quad\left. + \frac{x_{\ce{Si}}^{\mathrm{m}}}{2(1 - x_{\ce{Si}}^{\mathrm{m}})^2} - 1 \right) \nonumber,
\end{align}
\noindent and
\begin{align}
\ln \gamma_{\mathrm{O}}^{\mathrm{m}} = &4.29 - \frac{16,500}{T_{\mathrm{SMI}}~[\mathrm{K}]} + \frac{16,500}{T_{\mathrm{SMI}}~[\mathrm{K}]} \ln(1 - x_{\ce{O}}^{\mathrm{m}}) \\
&+ 5 \frac{1,873}{T_{\mathrm{SMI}}~[\mathrm{K}]} x_{\ce{Si}}^{\mathrm{m}} \left( 1 + \frac{\ln(1 - x_{\ce{Si}}^{\mathrm{m}})}{x_{\ce{Si}}^{\mathrm{m}}} - \frac{1}{1 - x_{\ce{O}}^{\mathrm{m}}} \right) \nonumber \\
&- 5 \frac{1,873}{T_{\mathrm{SMI}}~[\mathrm{K}]} (x_{\ce{Si}}^{\mathrm{m}})^2 x_{\ce{O}} \left( \frac{1}{1 - x_{\ce{O}}^{\mathrm{m}}} + \frac{1}{1 - x_{\ce{Si}}^{\mathrm{m}}} \right. \nonumber \\
&\quad\left. + \frac{x_{\ce{O}}^{\mathrm{m}}}{2(1 - x_{\ce{O}}^{\mathrm{m}})^2} - 1 \right) \nonumber.
\end{align}

The activity coefficient of \ce{H} in metal ($\gamma_{\mathrm{H}}^{\mathrm{m}}$) is derived from a pseudo-ternary mixing model for H in molten Fe as described in \citet{young_earth_2023}, following the prescription of \citet{righter_activity_2020}:

\begin{align}
\ln \gamma_{\ce{H}}^{\mathrm{m}}
= &-3.8 \, \bigl(x_{\ce{Si}}^{\mathrm{m}} + x_{\ce{O}}^{\mathrm{m}}\bigr) \\
&\times \left( 1
+ \frac{\ln\!\left(1 - x_{\ce{Si}}^{\mathrm{m}} - x_{\ce{O}}^{\mathrm{m}}\right)}
       {x_{\ce{Si}}^{\mathrm{m}} + x_{\ce{O}}^{\mathrm{m}}}
- \frac{1}{1 - x_{\ce{H}}^{\mathrm{m}}} \right) \nonumber \\
&\quad+ 3.8 \, \bigl(x_{\ce{Si}}^{\mathrm{m}} + x_{\ce{O}}^{\mathrm{m}}\bigr)^{2} \,
x_{\ce{H}}^{\mathrm{m}} \nonumber \\
&\times \left( \frac{1}{1 - x_{\ce{H}}^{\mathrm{m}}}
+ \frac{1}{1 - x_{\ce{Si}}^{\mathrm{m}} - x_{\ce{O}}^{\mathrm{m}}}\right.\nonumber \\
&\quad\left.+ \frac{x_{\ce{H}}^{\mathrm{m}}}{2 (1 - x_{\ce{H}}^{\mathrm{m}})^{2}}
- 1 \right) \nonumber.
\end{align}

In contrast to \citet{werlen_atmospheric_2025,werlen_sub-neptunes_2025}, carbon partitioning into the metal phase is not included in this study. This choice is made to maintain a model setup directly comparable to that of \citet{young_earth_2023}, which serves as the reference framework for our proto–Earth calculations and allows the effects of non-ideality to be isolated and assessed more clearly.

\subsubsection{Thermodynamic Data}

Our thermodynamic data follow the values reported in the appendix of \citet{schlichting_chemical_2022}.  

\begin{figure}[ht]
    \centering
    \includegraphics{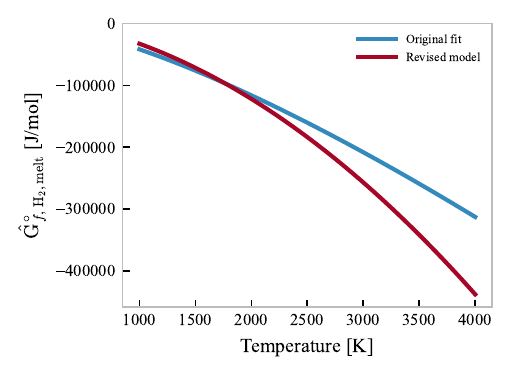}
    \caption{Gibbs free energy of \ce{H2} dissolution in silicate melt. The blue curve shows the \textit{original fit} based on the solubility experiments of \citet{hirschmann_solubility_2012}. The red curve shows the \textit{revised model}, which adopts the free energy of mixing from \citet{gilmore_core-envelope_2026} anchored to the experimental free energy of reaction from \citet{hirschmann_solubility_2012}. The revised relation yields more negative Gibbs free energies at high temperatures, implying greater \ce{H2} solubility in silicates at those conditions.}
    \label{fig:new_G_meltH2}
\end{figure}

The standard-state free energy of H$_2$ in melt is of particular interest.  We use a  revised standard-state molar Gibbs free energy for \ce{H2} in silicate melt. We derive this new value for $\hat{G}^{\circ}_{\rm H_2, melt}$ by combining the non-ideal free energy of mixing between \ce{MgSiO3} and \ce{H2} \citep{gilmore_core-envelope_2026}, the non-ideal EoS for H$_2$ \citep{Chabrier2019}, and  experimental constraints on H$_2$ solubility \citep{hirschmann_solubility_2012}. We solve the equation for the Gibbs free energy for the melt:

\begin{equation}
\begin{aligned}
\hat{G}_{\rm melt} =& x_{\rm H_2} \hat{G}^{\circ}_{f,{\rm H_2, melt}} \\ &+ x_{\rm MgSiO_3} \hat{G}^{\circ}_{f,{\rm MgSiO_3, melt}} +\Delta \hat{G}_{\rm mix}
\label{eqn:Gmelt}
\end{aligned}
\end{equation}

\noindent simultaneously with the equation relating the chemical potential of H$_2$ to that for H$_2$ in the melt at equilibrium: 

\begin{equation}
\begin{aligned}
\mu_{\rm H_2, gas}=\hat{G}^{\circ}_{f,{\rm H_2, melt}}  + \Delta \hat{G}_{\rm mix} + (1-x_{\rm H_2})\frac{\partial \hat{G}_{\rm mix}}{\partial x_{\rm H_2}},
\end{aligned}
\label{eqn:GH2melt}
\end{equation}
\noindent where the right-hand side of Equation \ref{eqn:GH2melt} is the chemical potential for H$_2$ in the melt and the chemical potential for the gas includes the non-ideality for H$_2$ (Equation \ref{eq:conditions}).  The two unknown variables of interest are $\hat{G}^{\circ}_{f,{\rm H_2, melt}}$ and $\hat{G}_{\rm melt}$ where the $^{\circ}$ symbol indicates the appropriate standard state. We solve the equations at a pressure of 0.17 GPa, thus anchoring our values for $\hat{G}^{\circ}_{f,{\rm H_2, melt}}$ to the values implied by the measured equilibrium constant for the H$_2$ solubility reaction at the experimental temperature used by \cite{hirschmann_solubility_2012} (i.e., $\hat{G}^{\circ}_{f,{\rm H_2, melt}}-\hat{G}^{\circ}_{f,{\rm H_2, gas}} = -RT\ln{k_{\rm eq}}$). We fit our results as a function of temperature where the $T$ dependence reflects the non-ideal mixing in the melt (Figure~\ref{fig:new_G_meltH2}). 

The effects of extreme pressures on free energies for melt species are not as well known as the temperature effects. While the effects of pressures of hundreds of GPa on molar volumes themselves are indeed large, the effects on equilibrium constants are comparatively small, as shown below. 

Correcting standard-state free energies for a reaction requires evaluating 

\begin{equation}
\Delta \ln k_{\rm eq}(P,T) = -\frac{1}{RT}\int_{P_0}^{P} \Delta \hat{V}_{\rm rxn}(P',T)\,dP',
\label{eqn:P_integral}
\end{equation}

\noindent where $\Delta \ln k_{\rm eq}(P,T)$ is the change in the log of the equilibrium constant due to pressure and $\Delta \hat{V}_{\rm rxn}$ is the molar change in volume for the reaction. The latter is obtained from the sum of molar volumes of the products minus those of the reactants, each weighted by their stoichiometric coefficients.  Detailed equations of state for each of the participating species are required to evaluate Equation \ref{eqn:P_integral}.  

In general, we do not yet have the requisite EoSs for all melt species.  Fortunately, the pressure effects at relevant pressure and temperatures can be small due to the fact that while molar volumes decrease significantly with pressure, their differences are less sensitive to pressure (i.e., they change similarly). We can use the intra-melt reaction 2 in our network (see Appendix~\ref{app:chemical_network}), $\rm MgSiO_3 \rightleftharpoons MgO + SiO_2$ (R2), as an example to illustrate the effects of pressure on our intra-melt equilibrium constants. \cite{schlichting_chemical_2022} point out that the opposing effects of pressure and temperature tend to cancel at the $T$ and $P$ conditions in the interiors of sub-Neptunes. We have estimates for the equations of state for all three components in R2, allowing us to quantify this effect. We present two approaches for evaluating pressure effects on intramelt reactions where full details of the EoSs remain uncertain.  The two approaches differ in the treatment of thermal pressure and the assumptions required.

In the first, simpler approach, the molar volumes of the three components are calculated using a fixed bulk modulus in evaluating thermal pressures. For each species in the reaction we adopt the third-order Birch - Murnaghan equation of state,
\begin{equation}
\begin{aligned}
P(V) =& \frac{3}{2}K_0\left[\left(\frac{V_0}{V}\right)^{7/3} - \left(\frac{V_0}{V}\right)^{5/3}\right] \\
&\left\{1 + \frac{3}{4}(K_0'-4)\left[\left(\frac{V_0}{V}\right)^{2/3} -1 \right]\right\},
\label{eqn:BM3}
\end{aligned}
\end{equation}
where $K_0$ is the bulk modulus at zero pressure, $K_0'$ its pressure derivative, and $V_0$ the reference volume at some reference temperature.

Thermal effects are estimated by defining the thermal pressure in terms of a pressure-dependent expansivity $\alpha(P)$:
\begin{equation}
P_{\mathrm{thermal}}(P,T) = \alpha(P)\,K_0\,(T - T_{\mathrm{ref}}), 
\label{eqn:Pthermal_simple}
\end{equation}
where $T_{\rm ref}$ refers to the temperature at which the EoS parameters are reported. Values for $\alpha(P)$ can be obtained using the Anderson - Gr\"{u}neisen relation \citep{Anderson_1966},
\begin{equation}
\alpha(P) = \alpha_0 \left[1 + \frac{K_0'}{K_0}P\right]^{-\delta_T/K_0'},
\end{equation}
where $\alpha_0$ is the reference expansivity and parameter $\delta_T = (\partial\ln{\alpha}/\partial\ln{V})_T$ is the Anderson-Gr\"{u}neisen parameter.  The latter has values of $\sim 2\gamma_g$ \citep{Chang_1967}, where $\gamma_g$ is the Gr\"{u}neisen parameter. For the present case,  values for $\gamma_g$ are about 1 to 1.25 for \ce{MgO} in the melt, 0.6 to 1 for melt \ce{MgSiO3}, and  0.1 to 1 for melt \ce{SiO2} at the pressures and temperatures of interest \citep{DeKoker2009}. We use this formulation to exclude thermal pressure effects when evaluating  volumes in Equation \ref{eqn:BM3}, as is appropriate when using an effective "cold" pressure, such that

\begin{equation}
\begin{aligned}
P_{\mathrm{eff}} = P - P_{\mathrm{thermal}}.
\label{eqn:Peff}
\end{aligned}
\end{equation}

\begin{figure*}
{\includegraphics[width=1\textwidth]{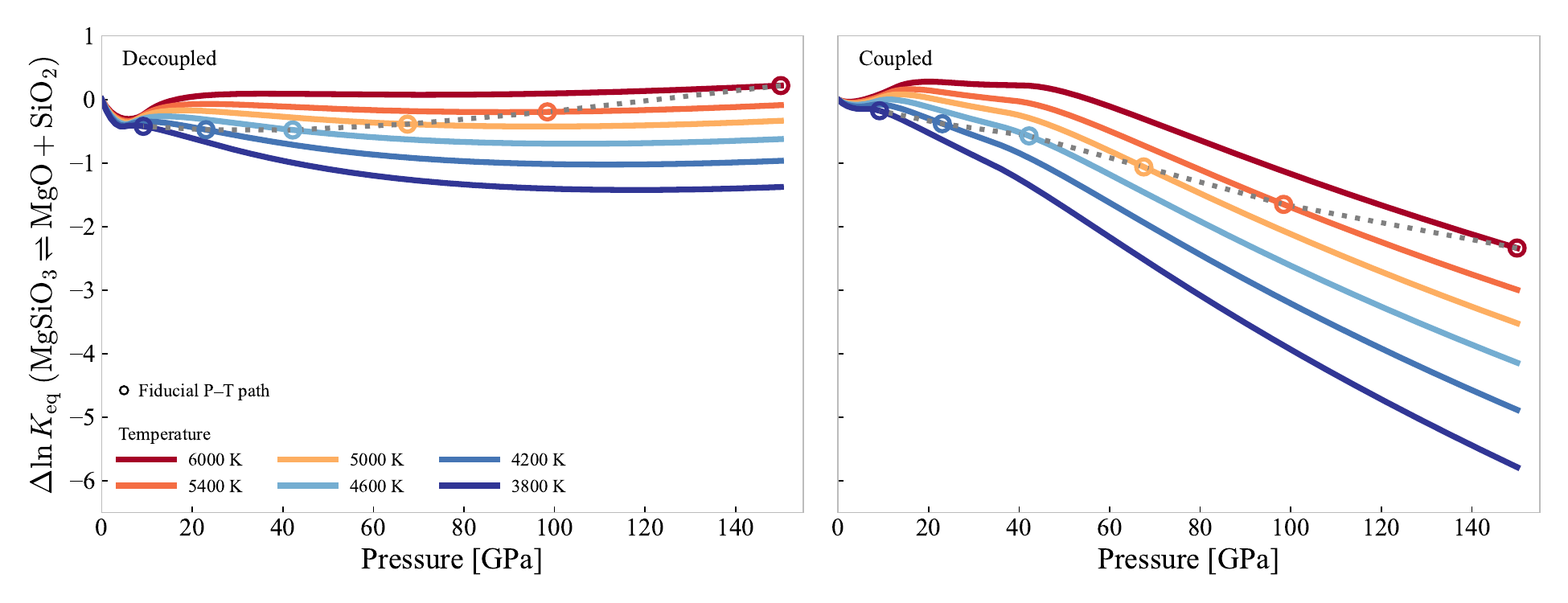}}
\caption{Plots of pressure corrections for the intra-melt reaction $\rm MgSiO_3 \rightleftharpoons MgO + SiO_2$ (R2) at temperatures ranging from 3800 K to 6000 K. Results using the simpler approach for estimating the effects of thermal pressure (Equation \ref{eqn:Pthermal_simple}) are shown in the left panel, and results using the integrated approach for estimating thermal pressure effects (Equation \ref{eqn:Pthermal_integrated}) are  shown at right.  A fiducial adiabat through a molten interior of a sub-Neptune with a total mass of 6 M$_\oplus$ and 3 weight percent hydrogen is shown by the open circles. See text.}
\label{fig:molarvolumes}
\end{figure*}

\noindent For this purpose we use the bulk moduli, expansivities, and Gr\"{u}neisen parameters for these melt species given by \cite{DeKoker2009}.
At each pressure and temperature, the molar volumes are obtained by solving Equation \ref{eqn:BM3} for each species based on the effective (cold) pressure, allowing implementation of Equation \ref{eqn:P_integral} to evaluate the pressure effects on the equilibrium constant for R2. The results are shown in Figure \ref{fig:molarvolumes} where the effects of pressure are seen to be relatively modest.

However, the first approach outlined above does not fully capture important temperature effects on volumes under compression. This is because in this decoupled formulation, thermal pressure is evaluated using the reference bulk modulus $K_0$ and a pressure-dependent expansivity, with temperature influencing the molar volumes only through the difference between the effective pressure and the total pressure, $P_{\mathrm{eff}} = P - P_{\mathrm{thermal}}$. Because elastic stiffening with compression does not feed back into the thermal pressure term, this formulation tends to underestimate the total thermal pressure at high pressure relative to more fully coupled treatments. At the same time, while we allow the thermal expansivity $\alpha$ to decrease monotonically with pressure using an Anderson--Gr\"uneisen parameterization, thermal pressure ultimately arises from anharmonic vibrational effects that are progressively suppressed under compression in a non-linear fashion. As a result, while $\alpha$ decreases with increasing pressure even as the bulk modulus $K$ increases due to elastic stiffening, the omission of non-linear effects associated with the suppression of anharmonic vibrational contributions means that the product $\alpha K = (\partial P/\partial T)_V$ may still be overestimated at high compression.

In order to better account for both $T$ and $P$ effects on volumes, thermal pressure may be treated self-consistently under compression by allowing the molar volumes to respond directly to temperature at fixed external pressure. This is possible given some additional inferences. In this approach we implement a coupled formulation in which the molar volume is obtained by solving implicitly for $V(P,T)$. Rather than evaluating the Birch-Murnaghan reference isotherm at an effective pressure that neglects feedbacks between thermal pressure and volumes, we instead explicitly enforce the condition

\begin{equation}
P_{\mathrm{BM3}}(V) + P_{\mathrm{thermal}}(V,T) = P,
\label{eq:pressure_balance}
\end{equation}

\noindent where the thermal pressure is expressed in terms of a pressure-dependent expansivity and the volume-dependent bulk modulus along the Birch-Murnaghan reference isotherm. This formulation allows thermal expansion and compression to interact explicitly, yielding a more realistic description of thermal pressure effects at high temperature and pressure; we solve directly for the volume at each $P$ and $T$.

Here we use $P_{\mathrm{BM3}}(V)$ to denote the pressure obtained from the third-order Birch-Murnaghan equation of state evaluated along a reference isotherm, which serves as the non-thermal baseline for the volume calculations. Correspondingly, $K_{\mathrm{BM3}}(V)$ denotes the isothermal bulk modulus derived from this reference Birch-Murnaghan relation.

In general, the thermal pressure of a condensed phase can be expressed using the Mie-Gr\"uneisen formulation:

\begin{equation}
P_{\mathrm{thermal}}(V,T) = \frac{\gamma(V)}{V}\,E_{\mathrm{thermal}}(V,T),
\end{equation}

\noindent where $\gamma(V)$ is the Gr\"uneisen parameter and $E_{\mathrm{thermal}}$ is the thermal internal energy relative to a reference isotherm. In the absence of explicit $\gamma(V)$ and heat capacity data for high-temperature, high-pressure melts, one can adopt a first-order linear expansion in terms of temperature, yielding

\begin{equation}
P_{\mathrm{thermal}}(V,T) \approx \alpha(V)\,K_{\mathrm{BM3}}(V)\,(T-T_{\mathrm{ref}}),
\label{eqn:Pthermal_integrated}
\end{equation}

\noindent that provides a closed-form, first-order approximation to the Mie-Gr\"uneisen thermal pressure in the absence of a complete thermodynamic equation of state (c.f., Equation \ref{eqn:Pthermal_simple}). In general, the thermal pressure coefficient may be written as $\alpha K_T$. In the coupled formulation adopted here, this coefficient is evaluated self-consistently along the Birch-Murnaghan reference isotherm, with the isothermal bulk modulus $K_T$ taken to be the volume-dependent modulus $K_{\mathrm{BM3}}(V)$ derived from the Birch-Murnaghan equation of state. The thermal expansivity $\alpha$ is evaluated at the compression state corresponding to the trial volume, equivalently expressed as a function of the non-thermal Birch-Murnaghan pressure $P_{\mathrm{BM3}}(V)$. Thermal pressure therefore represents the incremental pressure arising from heating relative to the reference temperature $T_{\mathrm{ref}}$ of the Birch-Murnaghan equation of state.

Here, the isothermal bulk modulus is obtained directly from the Birch--Murnaghan reference isotherm

\begin{equation}
K_{\mathrm{BM3}}(V)
=
- V\left(\frac{\partial P_{\mathrm{BM3}}}{\partial V}\right).
\end{equation}

The pressure dependence of the expansivity is represented using an Anderson--Gr\"uneisen-like exponential form,

\begin{equation}
\alpha(P_{\mathrm{BM3}})
=
\alpha_0 \exp\left[-\left(\frac{\delta_T}{K_T}\right)P_{\mathrm{BM3}}\right],
\end{equation}

\noindent where the isothermal bulk modulus $K_T$ is taken to be the reference bulk modulus $K_0$ for each species. This exponential form provides a useful and numerically stable parameterization of the Anderson--Gr\"uneisen relation when the bulk modulus varies with compression, while preserving the expected monotonic decay of $\alpha$ with increasing pressure.

At high compression, the product $\alpha K_{\mathrm{BM3}}$, as a linear approximation to the Mie-Gr\"uneisen formulation, can lead to excessive thermal pressures. To guard against this, we  invoke the assumption that because the decoupled approach does not account for the continued suppression of anharmonicity beyond the reference state,  $\alpha_0 K_0$ is proportional to a realistic upper bound on $(\partial P/\partial T)_V$.  We therefore impose a smooth, compression-dependent ``saturation'' on the effective coefficient $\alpha K_T$, such that

\begin{align}
\left(\alpha K\right)_{\mathrm{eff}}(V)
&=
\min\left[
\alpha(P_{\mathrm{BM3}})\,K_{\mathrm{BM3}}(V),
\right. \nonumber\\
&\left.\qquad
s_{\mathrm{mult}}\,\alpha_0 K_0
\right].
\end{align}

In this formulation, the thermal pressure coefficient is constrained to remain within a physically plausible range at high compression. Rather than allowing the linearized $\alpha K_T$ term to grow without bound as elastic stiffening increases $K_T$, we cap its magnitude at a multiple of the reference value $\alpha_0K_0$. This reflects the expectation from first-principles melt equations of state that thermal pressure contributions flatten at high compression. The resulting prescription recovers standard thermal expansion behavior at moderate pressures while avoiding unrealistically large thermal pressure contributions at extreme compression. We adopt $s_{\mathrm{mult}} = 2.5$, consistent with ab initio constraints for relevant silicate melts \citep[e.g.,][]{DeKoker2009}.

For each pressure and temperature, molar volumes are obtained from the roots of Equation~\ref{eq:pressure_balance} with respect to $V$ by solving for the volume at which the sum of the Birch-Murnaghan reference pressure and the thermal pressure equals the imposed external pressure. These self-consistent volumes that include expansivity effects can be used directly in the pressure integral for $\Delta \ln K_{\mathrm{eq}}$ at the specified temperature.  Resulting curves are shown in Figure \ref{fig:molarvolumes}.

While the pressure corrections for intramelt reactions are highly uncertain for many applications, the curves in Figure \ref{fig:molarvolumes} demonstrate that for most planetary applications, the standard-state free energies of formation for the melt species may suffice in the absence of the necessary equations of state. This is because the pressure corrections are often small in comparison to the large variability among equilibrium constants. Roughly, most planetary applications correspond to changes in $| \ln{k_{\rm eq}}|$ of $< 2$ for R2, i.e., variations in $k_{\rm eq} < 10\times$, and in many cases, changes in $k_{\rm eq}$ are $ < 3\times$.  For example, we map onto Figure \ref{fig:molarvolumes} pressures and temperatures corresponding to an adiabat for the molten interior of a model sub-Neptune planet with a mass of 6 M$_\oplus$ and 3 weight percent hydrogen, using the model procedures described in detail by \cite{young_differentiation_2025}. One can see that, by using the self-consistent thermal pressure  approach, the maximum shift in the equilibrium constant for reaction R2 in this case is roughly a factor of 10$\times$ at 140 GPa and 6000 K (shift in $| \ln{k_{\rm eq}}|$ of $< 2$). As pressure increases, so does temperature along the adiabat, resulting in a leveling off of the shift in the equilibrium constant deeper in the magma ocean. Given that differences in the equilibrium constants among the various melt reactions are generally orders of magnitude, and that similar shifts would apply to all intramelt reactions, this factor of 10 may not manifest as a substantial change in results. Generally, lower temperatures occur at lower pressures in planetary interiors, so the pressure correction is even smaller, although the combination of $T$ and $P$ is not always favorable for minimizing pressure corrections (e.g., for a molten proto-Earth). We conclude from this analysis that until we have detailed EoSs for all melt species, it is prudent to make use of the low-pressure, standard-state free energies of formation for most intra-melt reactions. Otherwise, one would be mixing low and high-pressure values inappropriately.

\begin{figure}
{\includegraphics{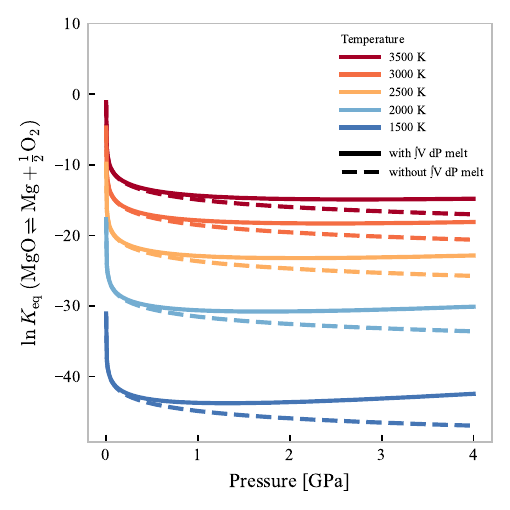}}\hfil
\caption{Plot of the log of the equilibrium constant for the reaction $\rm MgO_\text{silicate} \rightleftharpoons Mg_\text{gas} + 1/2O_{2\text{,gas}}$ as a function of pressure and temperature with and without pressure correction of the 1-bar standard-state Gibbs free energy of formation for the melt species \ce{MgO}, $\hat{G}^{o}_{f,{\rm MgO, melt}}$. Note the rapid initial decrease with $P$ is the same for both the corrected and uncorrected cases.}
\label{fig:MgO_volume}
\end{figure}

A similar argument holds for the effects of pressure corrections for melt species on the melt-rock-vapor reactions. We can use the same EoS for MgO melt to illustrate the effect of correcting the 1-bar Gibbs free energy for MgO melt to ambient pressure on the reaction $\rm MgO_\text{silicate} \rightleftharpoons Mg_\text{gas} + 1/2O_{2\text{,gas}}$. Here, we correct the Gibbs free energies of the gas-phase species for pressure, assuming ideal gas behavior. The corresponding equilibrium constant is

\begin{equation}
    k_{\rm eq}
    = \frac{x_{\text{Mg}}^{\text{g}}\,(x_{\text{O}_2}^{\text{g}})^{1/2}}{a_{\text{MgO}}^{\text{s}}}
    \left(\frac{P}{P^\circ}\right)^{3/2},
\end{equation}

\noindent where $P$ is the AMOI pressure and $P^\circ = 1$~bar is the standard-state reference pressure. Here $x_{i}^{j}$ denotes the mole fraction of species i in phase j, and $a_{\text{MgO}}^{\text{s}}$ is the activity of \ce{MgO} in silicate melt. Figure \ref{fig:MgO_volume} shows the effect of correcting $\hat{G}^{o}_{f,{\rm MgO, melt}}$ for pressure on $\ln(k_{\rm eq})$. The relative changes are small, especially at high temperatures. Here again, where the pressure effects on melt species are uncertain, consistent use of the more readily available low-P melt data will have minimal effects on the results.

Accordingly, all results presented below use 1-bar standard-state thermodynamic data for silicate and metal species, as pressure corrections to equilibrium constants are small over the pressure–temperature conditions explored here and because the necessary partial molar volume data are not available for all melt components.

\section{Results}\label{sec:Results}

Before turning to specific cases, we define the initial bulk composition adopted throughout this study. The planet is initialized with a silicate mass fraction of 65\% and a metal mass fraction of 35\%. The silicate mantle is composed of 94.7\% (mol) \ce{MgSiO3}, 3.3\% \ce{MgO}, 1.1\% \ce{SiO2}, 0.7\% \ce{Na2O}, 0.1\% \ce{Na2SiO3}, and minor amounts of \ce{FeO} and \ce{FeSiO3}. The initial envelope is \ce{H2}-dominated (99.9\% by mole), and the metal phase is 99.9\% by mole \ce{Fe}. This initial composition is used across all simulations, while the total planet mass and the envelope mass fraction are varied between the proto-Earth and sub-Neptune cases.

In total, we compare three model variants, labeled consistently with the figure legends: an ideal model (gas + silicate), in which gas and silicate species are treated ideally and non-ideality is included only for \ce{O} and \ce{Si} in the metal phase, consistent with the models of \citet{young_earth_2023}; a non-ideal ($\phi_i$) model, which incorporates non-ideal gas behavior via fugacity coefficients for \ce{H2}, \ce{H2O}, \ce{CH4}, \ce{CO2}, and \ce{CO}, together with non-ideal metal activities (\ce{O}, and \ce{Si}) while retaining ideal silicate activities; and a fully non-ideal ($\phi_i$,$\gamma_i$) model, which additionally includes non-ideal activity coefficients for the species \ce{H2O} and \ce{H2} in silicate melt and \ce{H} in the metal phase.

This three-model comparison is applied only to the sub-Neptune cases; for the proto-Earth case, we restrict the analysis to the ideal and fully non-ideal ($\phi_i$,$\gamma_i$) models.

\subsection{Proto-Earth as a Test Case}

\begin{figure*}
    \centering
    \includegraphics[width=1\textwidth]{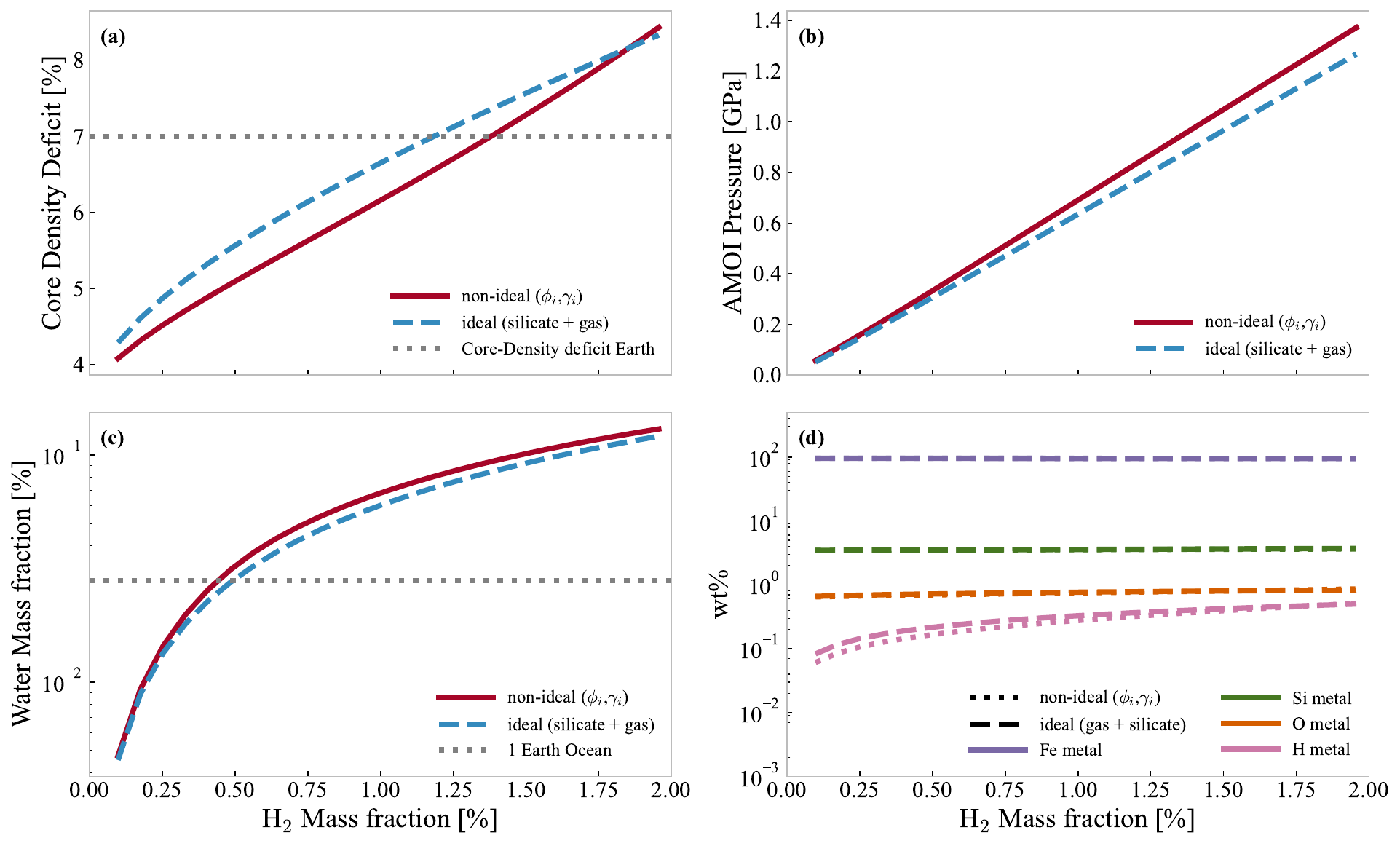}
    \caption{Results for a 0.5~M$_\oplus$ planetary embryo with an atmosphere–magma ocean interface (AMOI) temperature of 2350~K and a silicate–metal interface (SMI) temperature of 3000~K, shown as a function of the \ce{H2} mass fraction. Blue curves correspond to the ideal (gas + silicate) model, which includes non-ideal activity coefficients for \ce{O} and \ce{Si} in the metal phase. Red curves correspond to the fully non-ideal ($\phi_i$, $\gamma_i$) model, which additionally includes gas-phase fugacity coefficients for the gas species \ce{H2}, \ce{H2O}, \ce{CH4}, \ce{CO2}, and \ce{CO}, as well as non-ideal activity coefficients for \ce{H} in the metal phase and for \ce{H2O} and \ce{H2} in the silicate melt. Panel (a): core density deficit. For reference, the gray dotted line indicates Earth’s constraints on the core density deficit \citep{young_earth_2023}. Panel (b): AMOI pressure. Panel (c): water mass fraction, considering only \ce{H2O} in the gas and silicate melt. The gray dotted line indicates the water mass fraction equivalent to one Earth ocean, which serves as a lower bound on Earth’s total water budget. Panel (d): core composition (wt\% of metal species). Non-ideality slightly modifies the trends, but the overall behavior remains robust.}
    \label{fig:e_results}
\end{figure*}

We first examine how non-ideality affects the composition of a proto-Earth embryo of 0.5~M$_\oplus$, following \citet{young_earth_2023}. This choice of a sub-Earth mass embryo is motivated by standard models of terrestrial planet accretion \citep[e.g.,][]{chambers_planetary_2004,ricard_multi-phase_2009} and by previous models for the Moon-forming impact event involving global equilibration \citep[e.g.,][]{Canup_2012}.

We find that results are similar to those of \citet{young_earth_2023}. Specifically, equilibration with hydrogen simultaneously explains the density deficit of Earth's metal core and endogenous formation of ocean equivalents of water. This is noteworthy because \cite{young_earth_2023} used mainly ideal thermodynamics except for the metal phase. However, in the calculations the mass fraction of total hydrogen is about $1.3\%$ rather than the $\sim 0.2\%$ used in the previous study. The principal reason for this difference is the lower Gibbs free energy for H$_2$ in the silicate melt phase adopted here for both the ideal and non-ideal mixing calculations, as described above (Figure \ref{fig:new_G_meltH2}). We present results for an AMOI temperature of 2350~K and an SMI temperature of 3000~K for comparison with the original study. The total hydrogen content is varied up to 2~wt\%, which is roughly the maximum envelope mass fraction that a 0.5~M$_\oplus$ body can accrete \citep{ginzburg_super-earth_2016}.

Figure~\ref{fig:e_results} shows the results for the equilibrated embryo. Blue curves denote the ideal (gas + silicate) model, while red curves correspond to the fully non-ideal ($\phi_i$,$\gamma_i$) model. Panel (a) shows the uncompressed core density deficit, with the ideal treatment yielding a slightly larger deficit than the ideal case at the same mass fraction of H$_2$. Panel (b) displays the AMOI pressure, which shows no significant difference between the ideal and non-ideal case. The same is true for the global water mass fraction (panel c), calculated from \ce{H2O} in the gas and silicate phases as in \citet{werlen_sub-neptunes_2025}. Panel (d) shows the volatile content of the metal phase (wt\% of metal species); the non-ideal case exhibits a slightly lower hydrogen content, consistent with the lower core density deficit shown in Panel (a). The metal composition is basically the same as that in \cite{young_earth_2023}. Overall, non-ideality introduces only minor corrections, leaving the general trends in Earth-like embryos essentially unchanged.

\subsection{Sub-Neptunes}\label{sec:subneptunes_results}

We now examine the effect of non-ideal mixing on the AMOI pressure and volatile partitioning for a representative sub-Neptune with 4~M$_\oplus$, an AMOI temperature of 3000~K, and a SMI temperature of 3500~K.  

\subsubsection{Pressure and Water Content}
\begin{figure*}
    \centering
    \includegraphics[width=1\textwidth]{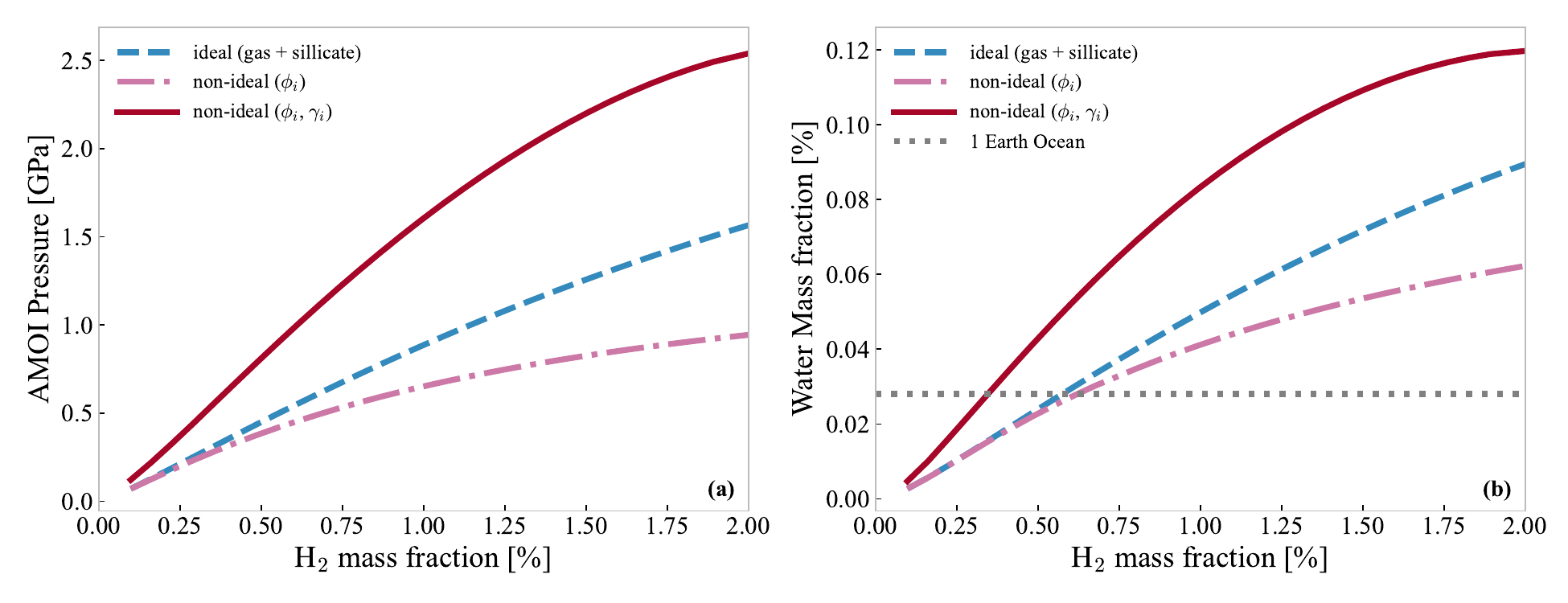}
    \caption{Results for a 4~M$_\oplus$ sub-Neptune with an atmosphere–magma ocean interface (AMOI) temperature of 3000~K and a silicate–metal equilibrium (SMI) temperature of 3500~K. The blue dashed line corresponds to the ideal (gas + silicate) model, in which gas and silicate species are treated ideally and non-ideal activity coefficients are included only for \ce{O} and \ce{Si} in the metal phase. The pink dashed–dotted line corresponds to the non-ideal ($\phi_i$) model, which additionally includes gas-phase fugacity coefficients for the gas species \ce{H2}, \ce{H2O}, \ce{CH4}, \ce{CO2}, and \ce{CO}. The red solid line corresponds to the fully non-ideal ($\phi_i$, $\gamma_i$) model, which further includes non-ideal activity coefficients for \ce{H2} and \ce{H2O} in the silicate melt and for \ce{H} in the metal phase. Panel (a): AMOI pressure. Panel (b): water mass fraction, considering only \ce{H2O} in the gas and silicate melt. The gray dotted line indicates the water mass fraction corresponding to one Earth ocean of water. Both pressure and water content decrease when only fugacity corrections are applied, but increase when activity corrections are added.}
    \label{fig:sN_P_H2O}
\end{figure*}

Figure~\ref{fig:sN_P_H2O} shows the AMOI pressure (panel a) and water mass fraction (panel b) as a function of the total hydrogen inventory. Non-ideality modifies both quantities in opposite directions. Accounting for fugacity coefficients ($\phi_i$) of the major gas species (\ce{H2}, \ce{H2O}, \ce{CH4}, \ce{CO2}, \ce{CO}) lowers the pressure and reduces the water content. This is because more hydrogen partitions into the melt, reducing the atmospheric mass fraction for the equilibrated planet. Introducing activity coefficients ($\gamma_i$) for silicate (\ce{H2}, \ce{H2O}) and metal (\ce{H}) instead suppresses hydrogen incorporation into the melt, increasing the atmospheric mass fraction and raising the pressure. The differences in water content mainly track these pressure variations, as the \ce{H2O} mole fractions in the gas and silicate remain similar across models.

\begin{figure*}
    \centering
    \includegraphics[width=1\textwidth]{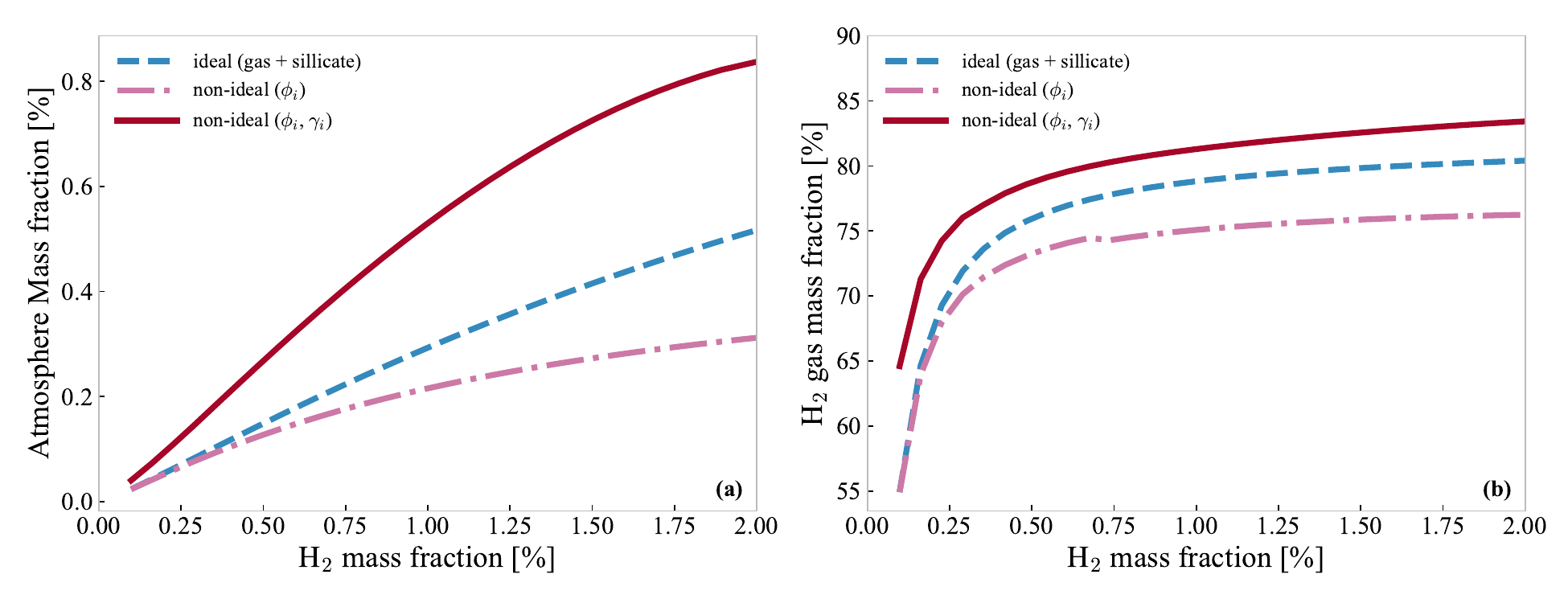}
    \caption{Same setup and color scheme as Figure~\ref{fig:sN_P_H2O}. Panel (a): atmosphere mass fraction. Panel (b): mass fraction of atmospheric \ce{H2}. Both quantities follow the same trends as pressure: fugacity corrections reduce both fractions, whereas additional activity corrections increase them above the ideal case.}    
    \label{fig:sN_mass fraction}
    \end{figure*}

\begin{figure}
    \centering
    \includegraphics{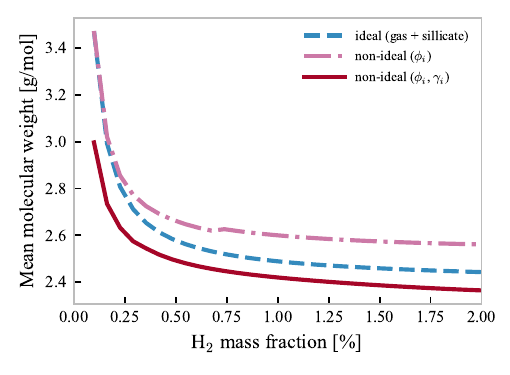}
    \caption{Same setup and  scheme as Figure~\ref{fig:sN_P_H2O}, but showing the mean molecular weight of the atmosphere. Atmospheres with higher H$_2$ fractions (fugacity-only case) have higher mean molecular weights, while including activity coefficients decreases the mean molecular weight.}
    \label{fig:mean_molecular_weight}
\end{figure}

\subsubsection{Atmosphere mass fraction and Mean Molecular Weight}

Beyond pressure and water content, the inclusion of non-ideality has further observational consequences. Figure~\ref{fig:sN_mass fraction} shows the atmosphere mass fraction together with the H$_2$ mass fraction of the atmosphere. The evolution of the atmosphere mass fraction closely tracks that of the pressure. Including only fugacity corrections lowers the atmosphere mass fraction, whereas the additional inclusion of activity coefficients for melt and metal species increases it again. The same trend is reflected in the H$_2$ mass fraction of the envelope. Since these atmospheres are H$_2$-dominated, the atmosphere mass fraction is indicative of the approximate hydrogen inventory stored in the interior. In the fugacity correction-only case, substantially more H$_2$ is sequestered into the interior compared to models that also include activity coefficients. Figure~\ref{fig:mean_molecular_weight} further illustrates that the mean molecular weight of the atmosphere is lower when only fugacity corrections are applied, consistent with the higher H$_2$ fractions in the envelope. Models including both fugacity and activity coefficients therefore predict atmospheres with higher mean molecular weights, which may be detectable.

\begin{figure}
    \centering
    \includegraphics{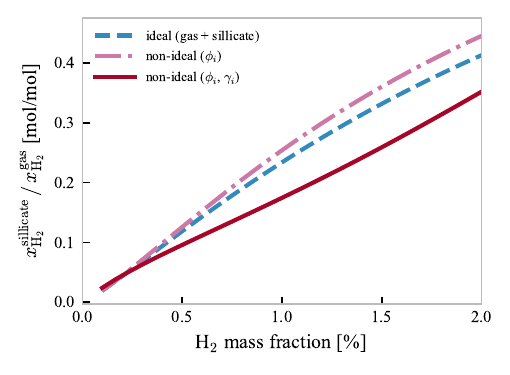}
    \caption{Same setup and color scheme as Figure~\ref{fig:sN_P_H2O}. Hydrogen solubility in the silicate, expressed as $x_{\ce{H2}}^{\mathrm{silicate}}/x_{\ce{H2}}^{\mathrm{gas}}$. Fugacity corrections enhance hydrogen solubility, while activity corrections reduce it.}
    \label{fig:sN_solubility_H2}
\end{figure}

\subsubsection{\ce{H2} Solubility as the Driver}

The differences in pressure, atmospheric mass fraction, and mean molecular weight (Figures~\ref{fig:sN_P_H2O}–\ref{fig:mean_molecular_weight}) can be traced back to the solubility of hydrogen. Figure~\ref{fig:sN_solubility_H2} shows that the solubility of hydrogen in silicates, expressed as $x_{\ce{H2}}^\text{silicate}/x_{\ce{H2}}^\text{gas}$, increases when gas-phase fugacity corrections are included but decreases when silicate activities are considered. Because hydrogen solubility rises steeply with AMOI temperature (Figure~\ref{fig:new_G_meltH2}), the magnitude of this effect depends strongly on the AMOI temperature. For Earth-sized embryos, the dissolved hydrogen mass fraction in silicate melt remains in a narrow range of 0.07–0.12 wt\% with increasing \ce{H2} mass fraction, meaning that non-ideal corrections shift the outcome only slightly. By contrast, sub-Neptunes host hotter surfaces and more massive envelopes, leading to dissolved hydrogen mass fractions spanning 0.06–1.45 wt\% as the atmospheric \ce{H2} inventory increases. In this regime, non-ideal corrections can amplify or suppress hydrogen uptake, producing deviations in atmospheric structure and composition. This explains why sub-Neptunes are more sensitive to non-ideality than terrestrial embryos and why a self-consistent treatment of non-ideality encompassing all phases is required to capture their volatile budgets.

\section{Discussion}\label{sec:Discussion}

Our results show that incorporating non-ideal behavior into equilibrium models requires treating the gas, melt, and metal chemistry. In the models investigated here, the silicate melt—through revised Gibbs free energies and activity coefficients of \ce{H} and \ce{H2O}—introduces effects that can offset those from gas-phase fugacity corrections.

These compensating effects are minor for Earth-sized embryos. For a 0.5~M$_\oplus$ proto-Earth, the AMOI pressure, global water mass fraction, and core density deficit remain nearly unchanged whether non-ideality is included or not. This indicates that the results of \citet{young_earth_2023}, who assumed ideal behavior in both the gas and melt phases, are not strongly affected by neglecting non-ideality. Their ability to reproduce key Earth properties—such as its water budget and core density deficit—through interactions between an \ce{H2}-rich envelope and a magma ocean therefore remains robust.

By contrast, the impact is somewhat larger for sub-Neptunes than for Earth-sized embryos. Their higher envelope mass fractions can yield hotter surface conditions, which in turn enhances \ce{H2} solubility in silicates. Under these circumstances, fugacity and activity corrections shift hydrogen partitioning and the balance between interior storage and atmospheric retention. Gas-phase fugacity coefficients increase hydrogen uptake by the melt, lowering the atmospheric mass fraction and AMOI pressure, while activity coefficients act in the opposite direction by suppressing melt solubility and restoring a more massive envelope. The resulting differences in pressure, water content, and atmospheric composition remain modest—typically less than 20\% and at most a factor of two—but are still noticeably larger than for terrestrial embryos. These shifts also translate into changes in the predicted mean molecular weight of the envelope, though again modest, at the ~10\% level.

We do not explicitly model the thermal and structural evolution of the planet in this study, but our results can be placed in context with previous work. In our models, the sub-Neptunes have envelope mass fractions of 0.2–0.8~wt\%. \citet{misener_importance_2022} showed that planets in this range cool and contract only on gigayear timescales. This suggests that global chemical equilibrium results presented here may capture the current state of many sub-Neptunes observed today.

\citet{bower_diversity_2025} incorporate solubilities and fugacity coefficients modeling volatile uptake through solubility laws and thereby treating the silicate melt as a chemically-inactive reservoir. Their framework does not include intra-melt reactions, atmosphere–melt exchange, or a metal phase. When we apply corrections for non-ideality in the gas phase only, our models reproduce their key result: fugacity coefficients lower AMOI pressures relative to the ideal case (Figure~\ref{fig:sN_P_H2O}). However, our inclusion of activity coefficients for selected melt and metal species counteracts this effect, leading instead to an overall increase in AMOI pressure. The implication is that non-ideality in both the gas and the melt phase should be considered, and that correcting for non-ideality in the gas phase without doing so in the melt phase may be worse than assuming ideality in both. 

Other equilibrium studies, such as \citet{tian_atmospheric_2024} and \citet{seo_role_2024}, include both fugacity and activity coefficients but with a smaller set of species and equations. Similar to \citet{bower_diversity_2025}, they also neglect the metal phase.

In a previous study, \citet{werlen_sub-neptunes_2025} analyzed the \ce{H2O} content of a synthetic population of sub-Neptunes and found that the maximum water mass fraction remains below 1.5~wt\%, well below the Hycean threshold of 10–90~wt\% \citep{madhusudhan_habitability_2021}. Our present results suggest that including non-ideality raises this upper limit, as activity corrections increase the water inventory. The effect is modest: the maximum water mass fraction rises only slightly, from 0.09 to 0.12~wt\% (an increase of about 25\% compared to the ideal case), and remains far from the Hycean regime.

\section{Conclusions}\label{sec:Conclusions}

In this study we coupled a global chemical equilibrium framework with non-ideal corrections to both the gas and melt phases. By incorporating fugacity coefficients for major volatile species and activity coefficients for selected silicate and metal components, we extended previous idealized approaches \citep[e.g.,][]{schlichting_chemical_2022,young_differentiation_2025,werlen_atmospheric_2025,werlen_sub-neptunes_2025} to capture more realistic mixing behavior under sub-Neptune conditions. This represents the first consistent treatment of non-ideality across all relevant phases in magma-ocean planets.

Our results demonstrate that non-ideality must be treated globally across all phases. Accounting for non-ideal effects in only one phase can lead to incomplete or even misleading trends. In our simulations, planetary embryos experience lower surface temperatures and hence reduced \ce{H2} solubility in silicates, so the influence of non-ideality is minimal. In these cases, the AMOI pressure, global water mass fraction, and core density deficit remain nearly unchanged whether fugacity and activity corrections are applied or not, explaining why idealized models still reproduce key Earth properties. By contrast, in the sub-Neptune case we find that including only fugacity corrections lowers the AMOI pressure, water mass fraction, atmosphere mass fraction, and \ce{H2} gas mass fraction relative to the ideal model, whereas adding both fugacity and activity coefficients reverses this trend, raising the quantities above the ideal case. The magnitude of these shifts remains modest—typically within 20\% and at most a factor of two—but they are nonetheless larger than for planetary embryos.

Our results further suggest that the so-called “fugacity crisis” \citep{Kite2019}—the apparent over-efficiency of hydrogen loss from the envelope into the interior implied by fugacity corrections alone—may be mitigated once melt activities are considered, which alter hydrogen partitioning across melt, metal, and gas.


\section*{Acknowledgments}

E.D.Y. acknowledges support from NASA grant number 80NSSC21K0477 issued through the Emerging Worlds program. H.E.S gratefully acknowledges support from NASA under grant number 80NSSC18K0828. C.D acknowledges support from the Swiss National Science Foundation under grant TMSGI2\_211313. A.S. acknowledges support from the Alfred P. Sloan Foundation AEThER grant G-2025-25284. This work has been carried out within the framework of the NCCR PlanetS supported by the Swiss National Science Foundation under grant 51NF40\_205606. We thank the anonymous reviewer for their insightful comments, which greatly helped to improve this study. We acknowledge the use of large language models (LLMs), including ChatGPT, to improve the grammar, clarity, and readability of the manuscript.

\section*{ORCID iDs}

\noindent 
Aaron Werlen \orcidlink{0009-0005-1133-7586} \href{https://orcid.org/0009-0005-1133-7586}{0009-0005-1133-7586} \\
Edward D. Young \orcidlink{0000-0002-1299-0801} 
\href{https://orcid.org/0000-0002-1299-0801}{0000-0002-1299-0801}\\
Hilke E. Schlichting \orcidlink{0000-0002-0298-8089} \href{https://orcid.org/0000-0002-0298-8089}{0000-0002-0298-8089} \\
Caroline Dorn \orcidlink{0000-0001-6110-4610} \href{https://orcid.org/0000-0001-6110-4610}{0000-0001-6110-4610} \\
Anat Shahar \orcidlink{0000-0002-0794-2717} \href{https://orcid.org/0000-0002-0794-2717}{0000-0002-0794-2717} \\

\appendix
\twocolumngrid

\section{Equilibrium Chemistry Methods}\label{app:equilibrium_methods}

Many studies adopt different notations and methods to solve equations defining the chemical thermodynamic equilibrium state of a system, potentially leading to some confusion.  Here, we place the method used here into context by comparing to two commonly used approaches: generalized Gibbs free energy minimization (GEM) and the application of the extended law of mass action (xLMA). We  show that the two approaches are conceptually equivalent.

We largely follow the notation and mathematical description from \citet{leal_overview_2017}. We refer the reader to this work for a comprehensive overview, including alternative numerical solution methods.

We define
\begin{equation}
    \vec{n} = [n_1,\dots,n_N]^T
\end{equation}
as the vector of molar amounts of each species comprising the phases of a system, where $n_i$ is the amount of the $i$th species. Likewise,
\begin{equation}
    \vec{b} = [b_1,\dots,b_M]^T
\end{equation}
denotes the vector of molar amounts of each element, where $b_j$ is the amount of the $j$th element. If $n$ or $b$ refers only to a specific phase (e.g., $\phi$), we write $n^\phi$ and $b^\phi$.  

The matrix $A$ is the coefficient matrix relating $b$ and $n$. Its $(j,i)$ entry is the number of moles of the $j$th element in the $i$th species. $A$ must be of full rank, meaning that all species are linearly independent.
  
As an example, consider  a system composed of a single gas phase with species \ce{H2O}, \ce{CH4}, \ce{CO2}, \ce{CO}, and \ce{O2}.  The vectors $n$ and $b$ are:
\begin{align}
    \vec{n} &= [n_{\ce{H2O}},\,n_{\ce{CH4}},\,n_{\ce{CO2}},\,n_{\ce{CO}},\,n_{\ce{O2}}]^T, \\
    \vec{b} &= [b_{\ce{H}},\,b_{\ce{C}},\,b_{\ce{O}}]^T.
\end{align}
The corresponding coefficient matrix $A$ is:
\begin{equation}\label{eq:example_A}
A =
\bordermatrix{
    & \ce{H2O} & \ce{CH4} & \ce{CO2} & \ce{CO} & \ce{O2} \cr
    \ce{H} & 2 & 4 & 0 & 0 & 0 \cr
    \ce{C} & 0 & 1 & 1 & 1 & 0 \cr
    \ce{O} & 1 & 0 & 2 & 1 & 2
}
\end{equation}
In general, $n_i$ will vary with varying conditions (temperature, pressure, and so forth) as prescribed by the minimum Legendre transform of internal energy, referred to as potentials, appropriate for the  variables of interest (e.g., Gibbs free energy where the relevant variables are $T$, $P$, and $n_i$, Helmnoltz free energy for $T$, $V$, and $n_i$, or enthalpy for $S$, $P$, and $n_k$) while $b_j$ will be fixed where the system is closed to mass transfer. 

\subsection{Gibbs Free Energy Minimization}

For a closed system where $\vec{b}$ is constant and at a specified temperature $T$ and pressure $P$, the species amounts $n_i$ must adopt specific values at chemical equilibrium in order to minimize the Gibbs free energy, $G$. The problem can be stated as:
\begin{equation}\label{ap_eq:equilibrium_cond}
    \min_{n \geq 0} G = \vec{n}^T \vec{\mu} \coloneq \sum_{i=1}^N n_i \mu_i,
\end{equation}
where $\vec{\mu } = [\mu_1,\dots,\mu_N]^T$ is the vector of chemical potentials of the species, defined as
\begin{equation}\label{ap_eq:chemical_potential}
    \mu_i = \mu_i^\circ + RT \ln a_i,
\end{equation}
with $\mu_i^\circ$ the standard-state chemical potential, $R$ the gas constant, and $a_i$ the activity of the $i$th species.

In addition to \autoref{ap_eq:equilibrium_cond}, mass must be conserved, which can be expressed as
\begin{equation}\label{ap_eq:mass_balance}
    A\vec{n} = \vec{b}.
\end{equation}

Equations \eqref{ap_eq:equilibrium_cond} and \eqref{ap_eq:mass_balance} can be transformed into a system of nonlinear equations via the Lagrangian:
\begin{equation}\label{eq:lagrangian}
    L(\vec{n},\vec{y}) = \vec{n}^T \vec{\mu } - \vec{y}^T (A\vec{n} - \vec{b}),
\end{equation}
where $y$ is the vector of Lagrange multipliers corresponding to the elemental mass balance constraint $A\vec{n} = \vec{b}$.

The necessary conditions for a local minimum at a stationary point of $L$ are:
\begin{align}\label{eq:system}
    \frac{\partial L}{\partial n} &= 0 \quad \text{if} \quad n_i > 0, \\
    \frac{\partial L}{\partial n} &\geq 0 \quad \text{if} \quad n_i = 0, \\
    \frac{\partial L}{\partial y} &= 0.
\end{align}

To remove the inequality, we introduce a slack variable $z_i$ defined by
\begin{equation}
    \frac{\partial L}{\partial n_i} = z_i.
\end{equation}
Using \autoref{eq:lagrangian}, the full system of non-linear equations, including mass conservation, becomes:
\begin{align}
    &\vec{\mu} - A^T \vec{y} - \vec{z} = 0, \label{eq:final_system}\\
    &A\vec{n} = \vec{b}, \\
    &n_i z_i = 0 \quad (i = 1,\dots,N), \\
    &n_i, z_i \geq 0 \quad (i = 1,\dots,N) \label{eq:final_system_2}.
\end{align}

This rigorous mathematical description can be readily transformed into the system of equations used in this study, as well as in \citet{schlichting_chemical_2022, young_earth_2023, werlen_sub-neptunes_2025, werlen_atmospheric_2025}.

To illustrate this, we use the same example as above. Assume that $n_i, z_i \geq 0$ and $n_i z_i = 0$, which means that all species are present — a condition that is always satisfied in the systems we consider. This implies $z_i = 0$ for all $i$. The condition for chemical equilibrium then reduces to:
\begin{align}
    &\vec{\mu} = A^T \vec{y},  \\
    &A\vec{n} = \vec{b}. \label{eq:mass_conservation_2}
\end{align}

When deriving chemical reactions, we identify the linearly independent reactions from the coefficient matrix, which is mathematically equivalent to finding the kernel $\vec{\nu}$ (null space) of the matrix:
\begin{equation}
    A \vec{\nu} = 0.
\end{equation}
This equation is equivalent to the chemical equilibrium condition. To show this, we multiply the chemical equilibrium condition by $\nu^T$:
\begin{equation}
    \vec{\nu}^T \vec{\mu} = \vec{\nu}^T A^T \vec{y} = (A \vec{\nu})^T \vec{y} = 0.
\end{equation}

\paragraph{Example.}  
For the example coefficient matrix given in \ref{eq:example_A}, the kernel is two-dimensional. Two example basis vectors of the kernel are:
\begin{align}
    \vec{\nu}_1 &= [2,\,-1,\,-3,\,4,\,0]^T, \\
    \vec{\nu}_2 &= [-2,\,1,\,-1,\,0,\,2]^T,
\end{align}
which correspond to the following two independent chemical reactions:
\begin{align}
    \ce{CH4 + 3CO2} &\rightleftharpoons \ce{2H2O + 4CO}, \\
    \ce{2H2O + CO2} &\rightleftharpoons \ce{CH4 + O2}.
\end{align}

Using the two example basis vectors, we obtain:
\begin{align}
    \vec{\nu}_1^T \vec{\mu} &= 2\mu_{\ce{H2O}} - \mu_{\ce{CH4}} - 3\mu_{\ce{CO2}} + 4\mu_{\ce{CO}} = 0, \\
    \vec{\nu}_2^T \vec{\mu} &= -2\mu_{\ce{H2O}} + \mu_{\ce{CH4}} - \mu_{\ce{CO2}} + 2\mu_{\ce{O2}} = 0.
\end{align}

Using $\mu_i = \mu_i^\circ + RT\ln a_i = G_i^\circ + RT\ln a_i$, where $G_i^\circ$ is the Gibbs free energy of species $i$ at standard state, these become:
\begin{equation}
    \begin{aligned}
    0 &= \Delta G_{\text{rxn,1}}^\circ\\ 
    &+ RT\left[ \ln a_{\ce{H2O}} - \ln a_{\ce{CH4}} - 3\ln a_{\ce{CO2}} + 4\ln a_{\ce{CO}} \right],
    \end{aligned}
\end{equation}
\begin{equation}
    \begin{aligned}
    0 &= \Delta G_{\text{rxn,2}}^\circ\\
    &+ RT\left[ -2\ln a_{\ce{H2O}} + \ln a_{\ce{CH4}} - \ln a_{\ce{CO2}} + 2\ln a_{\ce{O2}} \right],
    \end{aligned}
\end{equation}
where $\Delta G_{\text{rxn},j}^\circ$ is the Gibbs free energy of reaction $j$ at standard state.

The mass balance equations follow directly from \ref{eq:mass_conservation_2}. Using $n_i = x_i \cdot n_{\text{tot}}$ as well as $\sum_i x_i = 1$, we recover the equations used in this study.

\subsection{Extended Law of Mass-Action (xLMA) Equations}

It is possible to use the Lagrange multiplier $z_i$ as a stability criterion for a given phase. This is because $z_i > 0$ if species $n_i = 0$. The stability index of a phase $\Omega^\phi$ can be expressed as:
\begin{equation}
    \Omega^\phi \coloneq \sum_{k=1}^{N^\phi} x_k^\phi \exp\left(-\frac{z_k^\phi}{RT}\right),
\end{equation}
where $z_k^\phi$ is the stability index for the $k$th species in phase $\phi$ and $x_k^\phi$ is the mole fraction. If the phase is stable, $z_k^\phi = 0$ and $\Omega^\phi = 1$.

Consider the set of reactions
\begin{equation}
    0 \rightleftharpoons \sum_{i=1}^{N} \nu_{mi} \alpha_i \quad (m = 1,\dots,M),
\end{equation}
where $\alpha_i$ is the $i$th chemical species and $\nu_{mi}$ is the stoichiometric coefficient of the $m$th reaction for the $i$th species. We furthermore assume that the reactions are linearly independent.

The system is in equilibrium if the net reaction rates are zero, implying that all species are stable. Each reaction must therefore satisfy the standard law of mass action:
\begin{equation}\label{eq_ap:LMW}
    K_m = \prod_{i=1}^N a_i^{\nu_{mi}},
\end{equation}
where $K_m$ is the equilibrium constant of the $m$th reaction. The issue with \eqref{eq_ap:LMW} is that it is valid only at equilibrium; during the equilibrium calculation, some mass-action equations might need to be removed if certain phases are not stable.

The extended law of mass action (xLMA) resolves this by introducing a stability factor $w_i$:
\begin{equation}\label{eq_ap:equilibrium_constant}
    K_m = \prod_{i=1}^N (a_i w_i)^{\nu_{mi}},
\end{equation}
where
\begin{equation}
    w_i \coloneq \sum_{k=1}^{N^\phi} x_k^\phi w_k^\phi.
\end{equation}

By normalizing $z_i$ by $RT$, Equation~\eqref{eq_ap:equilibrium_constant} can be rewritten as:
\begin{equation}
    \ln K_m = \sum_{i=1}^N \nu_{mi} \left( \ln a_i - z_i \right).
\end{equation}

The resulting system of nonlinear equations which needs to be solved is:
\begin{align}
    &\ln \vec{K} = \nu \ln \vec{a} - \nu \vec{z},\label{eq:final_system_xLMA} \\
    &A\vec{n} = \vec{b}, \\
    &n_i z_i = 0 \quad (i = 1,\dots,N), \\
    &n_i, z_i \geq 0 \quad (i = 1,\dots,N) \label{eq:final_system_xLMA_2}.
\end{align}

Equations~\ref{eq:final_system_xLMA}-\ref{eq:final_system_xLMA_2} are mathematically identical to Equations~\ref{eq:final_system}-\ref{eq:final_system_2}. However, as noted by \citet{leal_overview_2017}, the xLMA formulation can offer computational advantages over direct Gibbs free energy minimization.

\section{Chemical Network}\label{app:chemical_network}

The chemical reaction network follows the formulation of \citet{schlichting_chemical_2022} and \citet{young_earth_2023}. In total, the system is described by 18 independent reactions involving 25 phase components. These encompass the silicate, metal, and gas reservoirs, with exchange allowed both within and between phases.

Below we present one possible basis set. Note that in equilibrium systems all linear combinations of a given basis set are, by definition, included in the reaction network \citep{schlichting_chemical_2022}.

Reactions within the silicate phase are:
\begin{equation}
\ce{Na2SiO3_{,silicate} \rightleftharpoons Na2O_{silicate} + SiO2_{,silicate}} \tag{R1}
\end{equation}
\begin{equation}
\ce{MgSiO3_{,silicate} \rightleftharpoons MgO_{silicate} + SiO2_{,silicate}} \tag{R2}
\end{equation}
\begin{equation}
\ce{FeO_{silicate} + 1/2Si_{metal} \rightleftharpoons Fe_{metal} + 1/2SiO2_{,silicate}} \tag{R3}
\end{equation}
\begin{equation}
\ce{FeSiO3_{,silicate} \rightleftharpoons FeO_{silicate} + SiO2_{,silicate}} \tag{R4}
\end{equation}

Reactions coupling the silicate and metal phases are:
\begin{equation}
\ce{O_{metal} + 1/2Si_{metal} \rightleftharpoons} \ce{1/2SiO2_{,silicate}} \tag{R5}
\end{equation}
\begin{equation}
\ce{2H_{metal} \rightleftharpoons H2_{,silicate}} \tag{R6}
\end{equation}
\begin{equation}
\ce{Si_{metal} + 2H2O_{silicate} \rightleftharpoons SiO2_{,silicate} + 2H2_{,silicate}} \tag{R7}
\end{equation}

Within the gas phase, the following reactions are included:
\begin{equation}
\ce{CO_{gas} + 1/2O2_{,gas} \rightleftharpoons CO2_{,gas}} \tag{R8}
\end{equation}
\begin{equation}
\ce{CH4_{,gas} + 1/2O2_{,gas} \rightleftharpoons} \ce{2H2_{,gas} + CO_{gas}} \tag{R9}
\end{equation}
\begin{equation}
\ce{H2_{,gas} + 1/2O2_{,gas} \rightleftharpoons H2O_{gas}} \tag{R10}
\end{equation}

Finally, magma ocean–atmosphere exchange is represented by:
\begin{equation}
\ce{FeO_{silicate} \rightleftharpoons Fe_{gas} + 1/2O2_{,gas}} \tag{R11}
\end{equation}
\begin{equation}
\ce{MgO_{silicate} \rightleftharpoons Mg_{gas} + 1/2O2_{,gas}} \tag{R12}
\end{equation}
\begin{equation}
\ce{SiO2_{,silicate} \rightleftharpoons SiO_{gas} + 1/2O2_{,gas}} \tag{R13}
\end{equation}
\begin{equation}
\ce{Na2O_{silicate} \rightleftharpoons} \ce{2Na_{gas} + 1/2O2_{,gas}} \tag{R14}
\end{equation}
\begin{equation}
\ce{H2_{,silicate} \rightleftharpoons H2_{,gas}} \tag{R15}
\end{equation}
\begin{equation}
\ce{H2O_{silicate} \rightleftharpoons H2O_{gas}} \tag{R16}
\end{equation}
\begin{equation}
\ce{CO_{silicate} \rightleftharpoons CO_{gas}} \tag{R17}
\end{equation}
\begin{equation}
\ce{CO2_{,silicate} \rightleftharpoons CO2_{,gas}} \tag{R18}
\end{equation}
\bibliography{references}

@article{chambers_planetary_2004,
	title = {Planetary accretion in the inner {Solar} {System}},
	volume = {223},
	issn = {0012-821X},
	url = {https://www.sciencedirect.com/science/article/pii/S0012821X04002791},
	doi = {10.1016/j.epsl.2004.04.031},
	number = {3},
	urldate = {2025-12-26},
	journal = {Earth and Planetary Science Letters},
	author = {Chambers, John E.},
	month = jul,
	year = {2004},
	pages = {241--252},
}

@article{Canup_2012,
author = {Robin M. Canup },
title = {Forming a Moon with an Earth-like Composition via a Giant Impact},
journal = {Science},
volume = {338},
number = {6110},
pages = {1052-1055},
year = {2012},
doi = {10.1126/science.1226073},
URL = {https://www.science.org/doi/abs/10.1126/science.1226073},
eprint = {https://www.science.org/doi/pdf/10.1126/science.1226073},
}

@article{ricard_multi-phase_2009,
	title = {A multi-phase model of runaway core–mantle segregation in planetary embryos},
	volume = {284},
	issn = {0012-821X},
	url = {https://www.sciencedirect.com/science/article/pii/S0012821X09002404},
	doi = {10.1016/j.epsl.2009.04.021},
	number = {1},
	urldate = {2025-12-26},
	journal = {Earth and Planetary Science Letters},
	author = {Ricard, Yanick and Šrámek, Ondřej and Dubuffet, Fabien},
	month = jun,
	year = {2009},
	pages = {144--150},
}

@article{rogers_redefining_2025,
	title = {Redefining interiors and envelopes: hydrogen–silicate miscibility and its consequences for the structure and evolution of sub-{Neptunes}},
	volume = {544},
	issn = {0035-8711},
	shorttitle = {Redefining interiors and envelopes},
	url = {https://doi.org/10.1093/mnras/staf1940},
	doi = {10.1093/mnras/staf1940},
	number = {4},
	urldate = {2025-12-21},
	journal = {Monthly Notices of the Royal Astronomical Society},
	author = {Rogers, James G and Young, Edward D and Schlichting, Hilke E},
	month = dec,
	year = {2025},
	pages = {3496--3511},
}

@article{felix_competing_2025,
	title = {Competing chemical signatures in the atmosphere of {TOI}-270 d: {Inference} of sulfur and carbon chemistry},
	volume = {701},
	copyright = {© The Authors 2025},
	issn = {0004-6361, 1432-0746},
	shorttitle = {Competing chemical signatures in the atmosphere of {TOI}-270 d},
	url = {https://www.aanda.org/articles/aa/abs/2025/09/aa55194-25/aa55194-25.html},
	doi = {10.1051/0004-6361/202555194},
	language = {en},
	urldate = {2025-12-21},
	journal = {Astronomy \& Astrophysics},
	author = {Felix, L. and Kitzmann, D. and Demory, B.-O. and Mordasini, C.},
	month = sep,
	year = {2025},
	note = {Publisher: EDP Sciences},
	pages = {A296},
}

@article{bower_diversity_2025,
	title = {Diversity of {Low}-mass {Planet} {Atmospheres} in the {C}–{H}–{O}–{N}–{S}–{Cl} {System} with {Interior} {Dissolution}, {Nonideality}, and {Condensation}: {Application} to {TRAPPIST}-1e and {Sub}-{Neptunes}},
	volume = {995},
	issn = {0004-637X},
	shorttitle = {Diversity of {Low}-mass {Planet} {Atmospheres} in the {C}–{H}–{O}–{N}–{S}–{Cl} {System} with {Interior} {Dissolution}, {Nonideality}, and {Condensation}},
	url = {https://doi.org/10.3847/1538-4357/ae1479},
	doi = {10.3847/1538-4357/ae1479},
	language = {en},
	number = {1},
	urldate = {2025-12-21},
	journal = {The Astrophysical Journal},
	author = {Bower, Dan J. and Thompson, Maggie A. and Hakim, Kaustubh and Tian, Meng and Sossi, Paolo A.},
	month = dec,
	year = {2025},
	note = {Publisher: The American Astronomical Society},
	pages = {59},
}

@article{lichtenberg_constraining_2025,
	title = {Constraining exoplanet interiors using observations of their atmospheres},
	volume = {390},
	url = {https://www.science.org/doi/10.1126/science.ads3360},
	doi = {10.1126/science.ads3360},
	number = {6769},
	urldate = {2025-12-21},
	journal = {Science},
	author = {Lichtenberg, Tim and Shorttle, Oliver and Teske, Johanna and Kempton, Eliza M.-R.},
	month = oct,
	year = {2025},
	note = {Publisher: American Association for the Advancement of Science},
	pages = {eads3660},
}

@article{nixon_magma_2025,
	title = {Magma {Ocean} {Interactions} {Can} {Explain} {JWST} {Observations} of the {Sub}-{Neptune} {TOI}-270 d},
	volume = {995},
	issn = {0004-637X},
	url = {https://doi.org/10.3847/1538-4357/ae17c8},
	doi = {10.3847/1538-4357/ae17c8},
	language = {en},
	number = {1},
	urldate = {2025-12-21},
	journal = {The Astrophysical Journal},
	author = {Nixon, Matthew C. and Somers, R. Sander and Savel, Arjun B. and Ih, Jegug and Kempton, Eliza M.-R. and Young, Edward D. and Schlichting, Hilke E. and Lichtenberg, Tim and Welbanks, Luis and Misener, William and Piette, Anjali A. A. and Wogan, Nicholas F.},
	month = dec,
	year = {2025},
	note = {Publisher: The American Astronomical Society},
	pages = {95},
}

@article{schaefer_chemistry_2009,
	title = {{CHEMISTRY} {OF} {SILICATE} {ATMOSPHERES} {OF} {EVAPORATING} {SUPER}-{EARTHS}},
	volume = {703},
	issn = {0004-637X},
	url = {https://doi.org/10.1088/0004-637X/703/2/L113},
	doi = {10.1088/0004-637X/703/2/L113},
	language = {en},
	number = {2},
	urldate = {2025-12-20},
	journal = {The Astrophysical Journal},
	author = {Schaefer, Laura and Fegley, Bruce},
	month = sep,
	year = {2009},
	note = {Publisher: The American Astronomical Society},
	pages = {L113},
}

@article{schaefer_outgassing_2007,
	title = {Outgassing of ordinary chondritic material and some of its implications for the chemistry of asteroids, planets, and satellites},
	volume = {186},
	copyright = {https://www.elsevier.com/tdm/userlicense/1.0/},
	issn = {00191035},
	url = {https://linkinghub.elsevier.com/retrieve/pii/S0019103506003174},
	doi = {10.1016/j.icarus.2006.09.002},
	language = {en},
	number = {2},
	urldate = {2025-12-20},
	journal = {Icarus},
	author = {Schaefer, Laura and Fegley, Bruce},
	month = feb,
	year = {2007},
	pages = {462--483},
}

@article{young_differentiation_2025,
	title = {Differentiation, the {Exception}, {Not} the {Rule}: {Evidence} for {Full} {Miscibility} in {Sub}-{Neptune} {Interiors}},
	volume = {6},
	issn = {2632-3338},
	shorttitle = {Differentiation, the {Exception}, {Not} the {Rule}},
	url = {https://iopscience.iop.org/article/10.3847/PSJ/ae1012/meta},
	doi = {10.3847/PSJ/ae1012},
	language = {en},
	number = {11},
	urldate = {2025-11-11},
	journal = {The Planetary Science Journal},
	author = {Young, Edward D. and Werlen, Aaron and Marcum, Sarah P. and Stixrude, Lars and Dullemond, Cornelis P.},
	month = nov,
	year = {2025},
	note = {Publisher: IOP Publishing},
	pages = {251},
}

@article{lee_mineral_2025,
	title = {Mineral {Cloud} {Formation} above {Magma} {Oceans} in {Sub}-{Neptune} {Atmospheres}},
	volume = {990},
	issn = {2041-8205},
	url = {https://doi.org/10.3847/2041-8213/adfe62},
	doi = {10.3847/2041-8213/adfe62},
	language = {en},
	number = {2},
	urldate = {2025-09-25},
	journal = {The Astrophysical Journal Letters},
	author = {Lee, Elspeth K. H. and Werlen, Aaron and Dorn, Caroline},
	month = sep,
	year = {2025},
	note = {Publisher: The American Astronomical Society},
	pages = {L43},
}

@article{werlen_sub-neptunes_2025,
	title = {Sub-{Neptunes} {Are} {Drier} than {They} {Seem}: {Rethinking} the {Origins} of {Water}-rich {Worlds}},
	volume = {991},
	issn = {2041-8205},
	shorttitle = {Sub-{Neptunes} {Are} {Drier} than {They} {Seem}},
	url = {https://dx.doi.org/10.3847/2041-8213/adff73},
	doi = {10.3847/2041-8213/adff73},
	language = {en},
	number = {1},
	urldate = {2025-09-19},
	journal = {The Astrophysical Journal Letters},
	author = {Werlen, Aaron and Dorn, Caroline and Burn, Remo and Schlichting, Hilke E. and Grimm, Simon L. and Young, Edward D.},
	month = sep,
	year = {2025},
	note = {Publisher: The American Astronomical Society},
	pages = {L16},
}

@article{gilmore_core-envelope_2026,
	title = {Core–envelope miscibility in sub-{Neptunes} and super-{Earths}},
	volume = {650},
	copyright = {2026 The Author(s)},
	issn = {1476-4687},
	url = {https://www.nature.com/articles/s41586-025-09970-4},
	doi = {10.1038/s41586-025-09970-4},
	language = {en},
	number = {8100},
	urldate = {2026-02-05},
	journal = {Nature},
	publisher = {Nature Publishing Group},
	author = {Gilmore, Travis and Stixrude, Lars},
	month = feb,
	year = {2026},
	pages = {60--64},
}

@article{righter_activity_2020,
	title = {Activity coefficients of siderophile elements in {Fe}-{Si} liquids at high pressure},
	issn = {24103403},
	url = {http://www.geochemicalperspectivesletters.org/article2034},
	doi = {10.7185/geochemlet.2034},
	language = {en},
	urldate = {2025-09-03},
	journal = {Geochemical Perspectives Letters},
	author = {Righter, K. and Rowland Ii, R. and Yang, S. and Humayun, M.},
	month = oct,
	year = {2020},
	pages = {44--49},
}

@article{hirschmann_solubility_2012,
	title = {Solubility of molecular hydrogen in silicate melts and consequences for volatile evolution of terrestrial planets},
	volume = {345-348},
	issn = {0012-821X},
	url = {https://www.sciencedirect.com/science/article/pii/S0012821X12003159},
	doi = {10.1016/j.epsl.2012.06.031},
	urldate = {2025-09-03},
	journal = {Earth and Planetary Science Letters},
	author = {Hirschmann, M. M. and Withers, A. C. and Ardia, P. and Foley, N. T.},
	month = sep,
	year = {2012},
	pages = {38--48},
}

@article{leal_overview_2017,
	title = {An overview of computational methods for chemical equilibrium and kinetic calculations for geochemical and reactive transport modeling},
	volume = {89},
	issn = {1365-3075, 0033-4545},
	url = {https://www.degruyter.com/document/doi/10.1515/pac-2016-1107/html},
	doi = {10.1515/pac-2016-1107},
	language = {en},
	number = {5},
	urldate = {2025-08-12},
	journal = {Pure and Applied Chemistry},
	author = {Leal, Allan M. M. and Kulik, Dmitrii A. and Smith, William R. and Saar, Martin O.},
	month = may,
	year = {2017},
	pages = {597--643},
}

@article{werlen_atmospheric_2025,
	title = {Atmospheric {C}/{O} {Ratios} of {Sub}-{Neptunes} with {Magma} {Oceans}: {Homemade} rather than {Inherited}},
	volume = {988},
	issn = {2041-8205},
	shorttitle = {Atmospheric {C}/{O} {Ratios} of {Sub}-{Neptunes} with {Magma} {Oceans}},
	url = {https://dx.doi.org/10.3847/2041-8213/adf185},
	doi = {10.3847/2041-8213/adf185},
	language = {en},
	number = {2},
	urldate = {2025-07-28},
	journal = {The Astrophysical Journal Letters},
	author = {Werlen, Aaron and Dorn, Caroline and Schlichting, Hilke E. and Grimm, Simon L. and Young, Edward D.},
	month = jul,
	year = {2025},
	note = {Publisher: The American Astronomical Society},
	pages = {L55},
}

@article{misener_atmospheres_2023,
	title = {Atmospheres as windows into sub-{Neptune} interiors: coupled chemistry and structure of hydrogen–silane–water envelopes},
	volume = {524},
	copyright = {https://academic.oup.com/journals/pages/open\_access/funder\_policies/chorus/standard\_publication\_model},
	issn = {0035-8711, 1365-2966},
	shorttitle = {Atmospheres as windows into sub-{Neptune} interiors},
	url = {https://academic.oup.com/mnras/article/524/1/981/7207406},
	doi = {10.1093/mnras/stad1910},
	language = {en},
	number = {1},
	urldate = {2025-07-17},
	journal = {Monthly Notices of the Royal Astronomical Society},
	author = {Misener, William and Schlichting, Hilke E and Young, Edward D},
	month = jul,
	year = {2023},
	note = {Publisher: Oxford University Press (OUP)},
	pages = {981--992},
}

@article{young_earth_2023,
	title = {Earth shaped by primordial {H2} atmospheres},
	volume = {616},
	copyright = {2023 The Author(s), under exclusive licence to Springer Nature Limited},
	issn = {1476-4687},
	url = {https://www.nature.com/articles/s41586-023-05823-0},
	doi = {10.1038/s41586-023-05823-0},
	language = {en},
	number = {7956},
	urldate = {2025-07-03},
	journal = {Nature},
	author = {Young, Edward D. and Shahar, Anat and Schlichting, Hilke E.},
	month = apr,
	year = {2023},
	note = {Publisher: Nature Publishing Group},
	pages = {306--311},
}

@article{badro_core_2015,
	title = {Core formation and core composition from coupled geochemical and geophysical constraints},
	volume = {112},
	url = {https://www.pnas.org/doi/10.1073/pnas.1505672112},
	doi = {10.1073/pnas.1505672112},
	number = {40},
	urldate = {2025-06-12},
	journal = {Proceedings of the National Academy of Sciences},
	author = {Badro, James and Brodholt, John P. and Piet, Hélène and Siebert, Julien and Ryerson, Frederick J.},
	month = oct,
	year = {2015},
	note = {Publisher: Proceedings of the National Academy of Sciences},
	pages = {12310--12314},
}

@article{misener_importance_2022,
	title = {The importance of silicate vapour in determining the structure, radii, and envelope mass fractions of sub-{Neptunes}},
	volume = {514},
	issn = {0035-8711},
	url = {https://ui.adsabs.harvard.edu/abs/2022MNRAS.514.6025M/abstract},
	doi = {10.1093/mnras/stac1732},
	language = {en},
	number = {4},
	urldate = {2025-05-15},
	journal = {Monthly Notices of the Royal Astronomical Society},
	author = {Misener, William and Schlichting, Hilke E.},
	month = aug,
	year = {2022},
	pages = {6025--6037},
}

@misc{burn_water-rich_2024,
	title = {Water-rich sub-{Neptunes} and rocky super {Earths} around different {Stars}: {Radii} shaped by {Volatile} {Partitioning}, {Formation}, and {Evolution}},
	shorttitle = {Water-rich sub-{Neptunes} and rocky super {Earths} around different {Stars}},
	url = {http://arxiv.org/abs/2411.16879},
	doi = {10.48550/arXiv.2411.16879},
	urldate = {2025-05-06},
	publisher = {arXiv},
	author = {Burn, Remo and Bali, Komal and Dorn, Caroline and Luque, Rafael and Grimm, Simon L.},
	month = nov,
	year = {2024},
	note = {arXiv:2411.16879 [astro-ph]},
}

@article{kite_water_2021,
	title = {Water on {Hot} {Rocky} {Exoplanets}},
	volume = {909},
	issn = {2041-8205, 2041-8213},
	url = {https://iopscience.iop.org/article/10.3847/2041-8213/abe7dc},
	doi = {10.3847/2041-8213/abe7dc},
	language = {en},
	number = {2},
	urldate = {2025-05-04},
	journal = {The Astrophysical Journal Letters},
	author = {Kite, Edwin S. and Schaefer, Laura},
	month = mar,
	year = {2021},
	pages = {L22},
}

@article{madhusudhan_new_2025,
	title = {New {Constraints} on {DMS} and {DMDS} in the {Atmosphere} of {K2}-18 b from {JWST} {MIRI}},
	volume = {983},
	issn = {2041-8205},
	url = {https://dx.doi.org/10.3847/2041-8213/adc1c8},
	doi = {10.3847/2041-8213/adc1c8},
	language = {en},
	number = {2},
	urldate = {2025-04-25},
	journal = {The Astrophysical Journal Letters},
	author = {Madhusudhan, Nikku and Constantinou, Savvas and Holmberg, Måns and Sarkar, Subhajit and Piette, Anjali A. A. and Moses, Julianne I.},
	month = apr,
	year = {2025},
	note = {Publisher: The American Astronomical Society},
	pages = {L40},
}

@article{madhusudhan_habitability_2021,
	title = {Habitability and {Biosignatures} of {Hycean} {Worlds}},
	volume = {918},
	issn = {0004-637X},
	url = {https://dx.doi.org/10.3847/1538-4357/abfd9c},
	doi = {10.3847/1538-4357/abfd9c},
	language = {en},
	number = {1},
	urldate = {2025-04-25},
	journal = {The Astrophysical Journal},
	author = {Madhusudhan, Nikku and Piette, Anjali A. A. and Constantinou, Savvas},
	month = aug,
	year = {2021},
	note = {Publisher: The American Astronomical Society},
	pages = {1},
}

@article{chachan_role_2018,
	title = {On the {Role} of {Dissolved} {Gases} in the {Atmosphere} {Retention} of {Low}-mass {Low}-density {Planets}},
	volume = {854},
	issn = {0004-637X},
	url = {https://dx.doi.org/10.3847/1538-4357/aaa459},
	doi = {10.3847/1538-4357/aaa459},
	language = {en},
	number = {1},
	urldate = {2025-04-09},
	journal = {The Astrophysical Journal},
	author = {Chachan, Yayaati and Stevenson, David J.},
	month = feb,
	year = {2018},
	note = {Publisher: The American Astronomical Society},
	pages = {21},
}

@article{tian_atmospheric_2024,
	title = {Atmospheric {Chemistry} of {Secondary} and {Hybrid} {Atmospheres} of {Super} {Earths} and {Sub}-{Neptunes}},
	volume = {963},
	issn = {0004-637X},
	url = {https://dx.doi.org/10.3847/1538-4357/ad217c},
	doi = {10.3847/1538-4357/ad217c},
	language = {en},
	number = {2},
	urldate = {2025-04-08},
	journal = {The Astrophysical Journal},
	author = {Tian, Meng and Heng, Kevin},
	month = mar,
	year = {2024},
	note = {Publisher: The American Astronomical Society},
	pages = {157},
}

@article{seo_role_2024,
	title = {Role of {Magma} {Oceans} in {Controlling} {Carbon} and {Oxygen} of {Sub}-{Neptune} {Atmospheres}},
	volume = {975},
	issn = {0004-637X, 1538-4357},
	url = {https://iopscience.iop.org/article/10.3847/1538-4357/ad7461},
	doi = {10.3847/1538-4357/ad7461},
	language = {en},
	number = {1},
	urldate = {2025-04-08},
	journal = {The Astrophysical Journal},
	author = {Seo, Chanoul and Ito, Yuichi and Fujii, Yuka},
	month = nov,
	year = {2024},
	pages = {14},
}

@article{young_phase_2024,
	title = {Phase {Equilibria} of {Sub}-{Neptunes} and {Super}-{Earths}},
	volume = {5},
	issn = {2632-3338},
	url = {https://iopscience.iop.org/article/10.3847/PSJ/ad8c40/meta},
	doi = {10.3847/PSJ/ad8c40},
	language = {en},
	number = {12},
	urldate = {2025-03-26},
	journal = {The Planetary Science Journal},
	author = {Young, Edward D. and Stixrude, Lars and Rogers, James G. and Schlichting, Hilke E. and Marcum, Sarah P.},
	month = dec,
	year = {2024},
	note = {Publisher: IOP Publishing},
	pages = {268},
}

@misc{benneke_jwst_2024,
	title = {{JWST} {Reveals} {CH4}, {CO2}, and {H2O} in a {Metal}-rich {Miscible} {Atmosphere} on a {Two}-{Earth}-{Radius} {Exoplanet}},
	url = {http://arxiv.org/abs/2403.03325},
	doi = {10.48550/arXiv.2403.03325},
	urldate = {2025-01-02},
	publisher = {arXiv},
	author = {Benneke, Björn and Roy, Pierre-Alexis and Coulombe, Louis-Philippe and Radica, Michael and Piaulet, Caroline and Ahrer, Eva-Maria and Pierrehumbert, Raymond and Krissansen-Totton, Joshua and Schlichting, Hilke E. and Hu, Renyu and Yang, Jeehyun and Christie, Duncan and Thorngren, Daniel and Young, Edward D. and Pelletier, Stefan and Knutson, Heather A. and Miguel, Yamila and Evans-Soma, Thomas M. and Dorn, Caroline and Gagnebin, Anna and Fortney, Jonathan J. and Komacek, Thaddeus and MacDonald, Ryan and Raul, Eshan and Cloutier, Ryan and Acuna, Lorena and Lafrenière, David and Cadieux, Charles and Doyon, René and Welbanks, Luis and Allart, Romain},
	month = mar,
	year = {2024},
	note = {arXiv:2403.03325 [astro-ph]},
}

@article{ginzburg_super-earth_2016,
	title = {Super-{Earth} {Atmospheres}: {Self}-consistent {Gas} {Accretion} and {Retention}},
	volume = {825},
	issn = {0004-637X},
	shorttitle = {Super-{Earth} {Atmospheres}},
	url = {https://ui.adsabs.harvard.edu/abs/2016ApJ...825...29G},
	doi = {10.3847/0004-637X/825/1/29},
	urldate = {2025-03-04},
	journal = {The Astrophysical Journal},
	author = {Ginzburg, Sivan and Schlichting, Hilke E. and Sari, Re'em},
	month = jul,
	year = {2016},
	note = {Publisher: IOP
ADS Bibcode: 2016ApJ...825...29G},
	pages = {29},
}

@article{kite_superabundance_2019,
	title = {Superabundance of {Exoplanet} {Sub}-{Neptunes} {Explained} by {Fugacity} {Crisis}},
	volume = {887},
	issn = {2041-8205},
	url = {https://dx.doi.org/10.3847/2041-8213/ab59d9},
	doi = {10.3847/2041-8213/ab59d9},
	language = {en},
	number = {2},
	urldate = {2024-12-28},
	journal = {The Astrophysical Journal Letters},
	author = {Kite, Edwin S. and Jr, Bruce Fegley and Schaefer, Laura and Ford, Eric B.},
	month = dec,
	year = {2019},
	note = {Publisher: The American Astronomical Society},
	pages = {L33},
}

@article{shorttle_distinguishing_2024,
	title = {Distinguishing oceans of water from magma on mini-{Neptune} {K2}-18b},
	volume = {962},
	issn = {2041-8205, 2041-8213},
	url = {http://arxiv.org/abs/2401.05864},
	doi = {10.3847/2041-8213/ad206e},
	language = {en},
	number = {1},
	urldate = {2024-08-22},
	journal = {The Astrophysical Journal Letters},
	author = {Shorttle, Oliver and Jordan, Sean and Nicholls, Harrison and Lichtenberg, Tim and Bower, Dan J.},
	month = feb,
	year = {2024},
	note = {arXiv:2401.05864 [astro-ph]},
	pages = {L8},
}

@article{madhusudhan_carbon-bearing_2023,
	title = {Carbon-bearing {Molecules} in a {Possible} {Hycean} {Atmosphere}},
	volume = {956},
	issn = {2041-8205},
	url = {https://dx.doi.org/10.3847/2041-8213/acf577},
	doi = {10.3847/2041-8213/acf577},
	language = {en},
	number = {1},
	urldate = {2024-05-05},
	journal = {The Astrophysical Journal Letters},
	author = {Madhusudhan, Nikku and Sarkar, Subhajit and Constantinou, Savvas and Holmberg, Måns and Piette, Anjali A. A. and Moses, Julianne I.},
	month = oct,
	year = {2023},
	note = {Publisher: The American Astronomical Society},
	pages = {L13},
}

@article{schlichting_chemical_2022,
	title = {Chemical {Equilibrium} between {Cores}, {Mantles}, and {Atmospheres} of {Super}-{Earths} and {Sub}-{Neptunes} and {Implications} for {Their} {Compositions}, {Interiors}, and {Evolution}},
	volume = {3},
	issn = {2632-3338},
	url = {https://iopscience.iop.org/article/10.3847/PSJ/ac68e6/meta},
	doi = {10.3847/PSJ/ac68e6},
	language = {en},
	number = {5},
	urldate = {2024-05-06},
	journal = {The Planetary Science Journal},
	author = {Schlichting, Hilke E and Young, Edward D.},
	month = may,
	year = {2022},
	note = {Publisher: IOP Publishing},
	pages = {127},
}

@article{Chabrier2019,
	doi = {10.3847/1538-4357/aaf99f},
	url = {https://doi.org/10.3847/1538-4357/aaf99f},
	year = 2019,
	publisher = {American Astronomical Society},
	volume = {872},
	number = {1},
	pages = {51},
	author = {G. Chabrier and S. Mazevet and F. Soubiran},
	title = {A New Equation of State for Dense Hydrogen{\textendash}Helium Mixtures},
	journal = {The Astrophysical Journal},
}

@article{Shi_Saxena_1992,
  author       = {Shi, Pingfang and Saxena, S K},
  title        = {Thermodynamic modeling of the C-H-O-S fluid system},
  url          = {https://www.osti.gov/biblio/7006422},
  journal      = {American Mineralogist; (United States)},
  issn         = {ISSN 0003-004X},
  volume       = {77:9-10},
  place        = {United States},
  year         = {1992},
  month        = {}}

@article{Haldemann,
	author = {Haldemann, Jonas and Alibert, Yann and Mordasini, Christoph and Benz, Willy},
	title = {AQUA: a collection of H2O equations of state for planetary models‚òÖ},
	DOI= "10.1051/0004-6361/202038367",
	url= "https://doi.org/10.1051/0004-6361/202038367",
	journal = {Astronomy \& Astrophysics},
	year = 2020,
	volume = 643,
	pages = "A105",
}

@article{Kerrick_Jacobs_1981,
  author = {Kerrick, D. M. and Jacobs, G. K.},
  title = {A modified Redlich-Kwong equation for H2O, CO2 and H2O-CO2 mixtures at elevated pressures and temperatures},
  journal = {American Journal of Science},
  volume = {281},
  pages = {735--767},
  year = {1981}
}

@ARTICLE{Ghiorso2002,
author = {{Ghiorso}, Mark S. and {Hirschmann}, Marc M. and {Reiners}, Peter W. and {Kress}, Victor C.},
title = "{The pMELTS: A revision of MELTS for improved calculation of phase relations and major element partitioning related to partial melting of the mantle to 3 GPa}",
journal = {Geochemistry, Geophysics, Geosystems},
year = 2002,
volume = {3},
number = {5},
eid = {1030},
pages = {1030},
doi = {10.1029/2001GC000217},
adsurl = {https://ui.adsabs.harvard.edu/abs/2002GGG.....3.1030G}}

@article{Makhluf_2017,
	Author = {Makhluf, A. R. and Newton, R. C. and Manning, C. E.},
	Da = {2017/09/01},
	Doi = {10.1134/S0869591117050046},
	Id = {Makhluf2017},
	Isbn = {1556-2085},
	Journal = {Petrology},
	Number = {5},
	Pages = {449--457},
	Title = {H2O activity in albite melts at deep crustal P-T conditions derived from melting experiments in the systems NaAlSi3O8-H2O-CO2 and NaAlSi3O8-H2O-NaCl},
	Ty = {JOUR},
	Url = {https://doi.org/10.1134/S0869591117050046},
	Volume = {25},
	Year = {2017}}

@article{Anderson_1966,
  title = {Derivation of Wachtman's Equation for the Temperature Dependence of Elastic Moduli of Oxide Compounds},
  author = {Anderson, Orson L.},
  journal = {Phys. Rev.},
  volume = {144},
  issue = {2},
  pages = {553--557},
  numpages = {0},
  year = {1966},
  month = {Apr},
  publisher = {American Physical Society},
  doi = {10.1103/PhysRev.144.553},
  url = {https://link.aps.org/doi/10.1103/PhysRev.144.553}
}

@article{Chang_1967,
title = {On the temperature dependence of the bulk modulus and the Anderson-Gr√ºneisen parameter Œ¥ of oxide compounds},
journal = {Journal of Physics and Chemistry of Solids},
volume = {28},
number = {4},
pages = {697-701},
year = {1967},
issn = {0022-3697},
doi = {https://doi.org/10.1016/0022-3697(67)90101-1},
url = {https://www.sciencedirect.com/science/article/pii/0022369767901011},
author = {Y.A. Chang}
}

@article{Kovacevic_2022,
	Author = {Kova{\v c}evi{\'c}, Tanja and Gonz{\'a}lez-Cataldo, Felipe and Stewart, Sarah T. and Militzer, Burkhard},
	Da = {2022/07/29},
	Doi = {10.1038/s41598-022-16816-w},
	Id = {Kova{\v c}evi{\'c}2022},
	Isbn = {2045-2322},
	Journal = {Scientific Reports},
	Number = {1},
	Pages = {13055},
	Title = {Miscibility of rock and ice in the interiors of water worlds},
	Ty = {JOUR},
	Url = {https://doi.org/10.1038/s41598-022-16816-w},
	Volume = {12},
	Year = {2022}}

@article{
Walker2022,
author = {David Walker  and Shuo Ding  and Yves Moussallam },
title = {Does f\uppercase{O}2 influence reversible silicate melting without redox-active cations?},
journal = {Proceedings of the National Academy of Sciences},
volume = {119},
number = {41},
pages = {e2211358119},
year = {2022},
doi = {10.1073/pnas.2211358119},
URL = {https://www.pnas.org/doi/abs/10.1073/pnas.2211358119},
eprint = {https://www.pnas.org/doi/pdf/10.1073/pnas.2211358119}}

@article{Henderson2006,
  title={The Structure of Silicate Glasses and Melts},
  author={Grant S. Henderson and Georges Calas and Jonathan F. Stebbins},
  journal={Elements},
  year={2006},
  volume={2},
  pages={269-273}
}

@article{Mysen1982,
author = {Mysen, Bjorn O. and Virgo, David and Seifert, Friedrich A.},
title = {The structure of silicate melts: Implications for chemical and physical properties of natural magma},
journal = {Reviews of Geophysics},
volume = {20},
number = {3},
pages = {353-383},
doi = {https://doi.org/10.1029/RG020i003p00353},
url = {https://agupubs.onlinelibrary.wiley.com/doi/abs/10.1029/RG020i003p00353},
eprint = {https://agupubs.onlinelibrary.wiley.com/doi/pdf/10.1029/RG020i003p0035},
year = {1982}
}

@article{Hess1971,
title = {Polymer model of silicate melts},
journal = {Geochimica et Cosmochimica Acta},
volume = {35},
number = {3},
pages = {289-306},
year = {1971},
issn = {0016-7037},
doi = {https://doi.org/10.1016/0016-7037(71)90038-X},
url = {https://www.sciencedirect.com/science/article/pii/001670377190038X},
author = {Paul C Hess},
}

@article{Nesbitt2020,
url = {https://doi.org/10.2138/am-2020-6841},
title = {Polymerization during melting of ortho- and meta-silicates: Effects on Q species stability, heats of fusion, and redox state of mid-ocean range basalts (MORBs)},
title = {},
author = {H. Wayne Nesbitt and G. Michael Bancroft and Grant S. Henderson},
pages = {716--726},
volume = {105},
number = {5},
journal = {American Mineralogist},
doi = {doi:10.2138/am-2020-6841},
year = {2020},
lastchecked = {2022-11-26}
}

@article{Ford1983,
    author = {Ford, C. E. and Russell, D. G. and Craven, J. A. and Fisk, M. R.},
    title = "{Olivine-Liquid Equilibria: Temperature, Pressure and Composition Dependence of the Crystal/Liquid Cation Partition Coefficients for Mg, Fe2+, Ca and Mn}",
    journal = {Journal of Petrology},
    volume = {24},
    number = {3},
    pages = {256-266},
    year = {1983},
    issn = {0022-3530},
    doi = {10.1093/petrology/24.3.256},
    url = {https://doi.org/10.1093/petrology/24.3.256},
    eprint = {https://academic.oup.com/petrology/article-pdf/24/3/256/4202351/24-3-256.pdf}
}

@article{Ryerson1985,
title = {Oxide solution mechanisms in silicate melts: Systematic variations in the activity coefficient of SiO2},
journal = {Geochimica et Cosmochimica Acta},
volume = {49},
number = {3},
pages = {637-649},
year = {1985},
issn = {0016-7037},
doi = {https://doi.org/10.1016/0016-7037(85)90159-0},
url = {https://www.sciencedirect.com/science/article/pii/0016703785901590},
author = {F.J Ryerson},
}

@article{Wood2013,
  title={Activities and volatilities of trace components in silicate melts: a novel use of metal–silicate partitioning data},
  author={Bernard J. Wood and Jon Wade},
  journal={Contributions to Mineralogy and Petrology},
  year={2013},
  volume={166},
  pages={911-921}
}

@article{Nicholls1971,
	Author = {Nicholls, J. and Carmichael, I. S. E. and Stormer, J. C.},
	Da = {1971/03/01},
	Doi = {10.1007/BF00373791},
	Id = {Nicholls1971},
	Isbn = {1432-0967},
	Journal = {Contributions to Mineralogy and Petrology},
	Number = {1},
	Pages = {1--20},
	Title = {Silica activity and \uppercase{P} total in igneous rocks},
	Ty = {JOUR},
	Url = {https://doi.org/10.1007/BF00373791},
	Volume = {33},
	Year = {1971}
 }

@article{Carmichael1970,
    author = {Carmichael, I. S. E. and Nicholls, J. and Smith, A. L.},
    title = "{Silica activity in igneous rocks}",
    journal = {American Mineralogist},
    volume = {55},
    number = {1-2},
    pages = {246-263},
    year = {1970},
    issn = {0003-004X},
    eprint = {https://pubs.geoscienceworld.org/msa/ammin/article-pdf/55/1-2/246/4251427/am-1970-246.pdf},
}

@article{Holzheid1997,
title = {The activities of \uppercase{N}i\uppercase{O}, \uppercase{C}o\uppercase{O} and \uppercase{F}e\uppercase{O} in silicate melts},
journal = {Chemical Geology},
volume = {139},
number = {1},
pages = {21-38},
year = {1997},
issn = {0009-2541},
doi = {https://doi.org/10.1016/S0009-2541(97)00030-2},
url = {https://www.sciencedirect.com/science/article/pii/S0009254197000302},
author = {A. Holzheid and H. Palme and S. Chakraborty},
}

@article{Hastie1986,
	Author = {J.W. Hastie and D.W. Bonnell},
	Doi = {https://doi.org/10.1016/0022-3093(86)90772-6},
	Issn = {0022-3093},
	Journal = {Journal of Non-Crystalline Solids},
	Number = {1},
	Pages = {151-158},
	Title = {{A predictive thermodynamic model of oxide and halide glass phase equilibria}},
	Url = {https://www.sciencedirect.com/science/article/pii/0022309386907726},
	Volume = {84},
	Year = {1986}}

@article{Corgne2008,
title = {Metal-silicate partitioning and constraints on core composition and oxygen fugacity during Earth accretion},
journal = {Geochimica et Cosmochimica Acta},
volume = {72},
number = {2},
pages = {574-589},
year = {2008},
issn = {0016-7037},
doi = {https://doi.org/10.1016/j.gca.2007.10.006},
url = {https://www.sciencedirect.com/science/article/pii/S0016703707005674},
author = {Alexandre Corgne and Shantanu Keshav and Bernard J. Wood and William F. McDonough and Yingwei Fei},
}

@article{DeKoker2009,
    author = {de Koker, Nico and Stixrude, Lars},
    title = "{Self-consistent thermodynamic description of silicate liquids, with application to shock melting of MgO periclase and MgSiO3 perovskite}",
    journal = {Geophysical Journal International},
    volume = {178},
    number = {1},
    pages = {162-179},
    year = {2009},
    issn = {0956-540X},
    doi = {10.1111/j.1365-246X.2009.04142.x},
    url = {https://doi.org/10.1111/j.1365-246X.2009.04142.x},
    eprint = {https://academic.oup.com/gji/article-pdf/178/1/162/5992596/178-1-162.pdf},
}

@ARTICLE{Kite_2020b,
        title={Atmosphere origins for exoplanet sub-neptunes},
  author={Kite, Edwin S and Fegley Jr, Bruce and Schaefer, Laura and Ford, Eric B},
  journal={The Astrophysical Journal},
  volume={891},
  number={2},
  pages={111},
  year={2020},
  publisher={IOP Publishing}
}

@article{Fegley1987,
	Author = {{Fegley}, Bruce, Jr. and {Cameran}, A. G. W.},
	Journal = {Earth and Planetary Science Letters},
	Pages = {207-222},
	Title = {A vaporization model for iron/silicate fractionation in the Mercury protoplanet},
	Volume = {82},
	Year = {1987}}

@ARTICLE{Kite2019,
       author = {{Kite}, Edwin S. and {Fegley}, Bruce, Jr. and {Schaefer}, Laura and {Ford}, Eric B.},
        title = "{Superabundance of Exoplanet Sub-Neptunes Explained by Fugacity Crisis}",
      journal = {\apjl},
         year = 2019,
        month = dec,
       volume = {887},
       number = {2},
          eid = {L33},
        pages = {L33},
          doi = {10.3847/2041-8213/ab59d9},
archivePrefix = {arXiv},
       eprint = {1912.02701},
 primaryClass = {astro-ph.EP},
       adsurl = {https://ui.adsabs.harvard.edu/abs/2019ApJ...887L..33K}
}
\bibliographystyle{aasjournal}



\end{document}